\documentclass[review]{elsarticle}

\usepackage{lineno,hyperref}

\usepackage{amsmath,amsthm,amsfonts,bm}	 
\usepackage{subfigure}
\usepackage{float}
\usepackage{multirow}
\usepackage{adjustbox}

\journal{International Journal of Heat and Mass Transfer}

\begin{document}

\begin{frontmatter}

\title{Particle-resolved thermal lattice Boltzmann simulation using OpenACC on multi-GPUs
\footnote{\href{DOI: 10.1016/j.ijheatmasstransfer.2023.124758}{DOI: DOI: 10.1016/j.ijheatmasstransfer.2023.124758}}
\footnote{\copyright \ 2023. This manuscript version is made available under the CC-BY-NC-ND 4.0 license http://creativecommons.org/licenses/by-nc-nd/4.0/}
}

\author[mymainaddress,mysecondaryaddress,mythirdaddress]{Ao Xu\corref{mycorrespondingauthor}}
\cortext[mycorrespondingauthor]{Corresponding author}
\ead{axu@nwpu.edu.cn}

\author[mymainaddress]{Bo-Tao Li}

\address[mymainaddress]{School of Aeronautics, Northwestern Polytechnical University, Xi’an 710072, China}
\address[mysecondaryaddress]{Institute of Extreme Mechanics, Northwestern Polytechnical University, Xi’an 710072, China}
\address[mythirdaddress]{Key Laboratory of Icing and Anti/De-icing, China Aerodynamics Research and Development Center, Mianyang 621000, China}

\begin{abstract}
We utilize the Open Accelerator (OpenACC) approach for graphics processing unit (GPU) accelerated particle-resolved thermal lattice Boltzmann (LB) simulation.
We adopt the momentum-exchange method to calculate fluid-particle interactions to preserve the simplicity of the LB method.
To address load imbalance issues, we extend the indirect addressing method to collect fluid-particle link information at each timestep and store indices of fluid-particle link in a fixed index array.
We simulate the sedimentation of 4,800 hot particles in cold fluids with a domain size of $4000^{2}$, and the simulation achieves 1750 million lattice updates per second (MLUPS) on a single GPU.
Furthermore, we implement a hybrid OpenACC and message passing interface (MPI) approach for multi-GPU accelerated simulation.
This approach incorporates four optimization strategies, including building domain lists, utilizing request-answer communication, overlapping communications with computations, and executing computation tasks concurrently.
By reducing data communication between GPUs, hiding communication latency through overlapping computation, and increasing the utilization of GPU resources, we achieve improved performance, reaching 10846 MLUPS using 8 GPUs.
Our results demonstrate that the OpenACC-based GPU acceleration is promising for particle-resolved thermal lattice Boltzmann simulation.
\end{abstract}


\end{frontmatter}


\section{Introduction}

Particle-laden thermal convection occurs ubiquitously in nature and daily life \cite{mathai2020bubbly,brandt2022particle}.
For example, under the action of a wind field, sand particles of different sizes follow the airflow with suspended or leaping motion, and these particles may form dust storms by unstable thermal conditions during strong winds \cite{wang2023drag}.
Another example is that under the influence of air conditioning and the ventilation system, atmospheric pollutant particles (PM10 and PM2.5) originating from dust and smoke can be suspended in the air for an extended period, and they may spread to a wider area with the aid of airflow \cite{norback2019sources}.

Numerical models for simulating particle-laden flows can be generally classified into two categories \cite{maxey2017simulation}:
the point-particle model and the particle-resolved model.
In the point-particle model, solid particles are treated as discrete masses that are much smaller than the mesh size of the computational grid. Empirical correlations, such as Stokes' viscous drag law, are used to calculate forces exerted on the solid particles by the fluid.
The advantage of the point-particle model is its relatively low computational load, making it suitable for tracking large numbers of particles (but with limited particle volume fractions) \cite{patovcka2020settling,patovcka2022residence,yang2022dynamic,yang2022energy}.
In contrast, in the particle-resolved model, particle sizes are larger than the resolution of the computational grid (known as finite-size particles).
They ensure the no-slip velocity boundary condition at the particle surface, and the forces and moments acting on the particles by the fluid are explicitly calculated by considering the interaction between the fluid phase and the solid phase.
The advantage of the particle-resolved model is its ability to accurately describe fluid-particle interactions based on first principles. Popular particle-resolved model includes the arbitrary Lagrangian-Eulerian method \cite{hu2001direct}, immersed boundary method \cite{peskin2002immersed}, fictitious domain method \cite{glowinski2001fictitious,yu2006fictitious}, and lattice Boltzmann (LB) method \cite{aidun2010lattice}.
The particle-resolve model can simulate dense suspension \cite{tao2018combined,walayat2018efficient,suzuki2020numerical,demou2022turbulent};
however, their high computational cost limits the number of particles that can be tracked.

Among these particle-resolved methods, the LB method is fascinating due to its ability to incorporate mesoscopic physical pictures while recovering macroscopic physical conservation laws with a relatively low computational cost.
Open-source codes based on the LB method, including OpenLB \cite{krause2021openlb}, Palabos \cite{latt2021palabos}, and Sailfish \cite{januszewski2014sailfish}, have facilitated the application of LB simulations in large-scale engineering problems  \cite{amati2021projecting}.
Over the past decades, the advancement of general-purpose graphics processing units (GPUs) has significantly improved high-performance computing, enabling faster simulations of larger physical domains or higher computational resolutions \cite{liu2019sunwaylb,falcucci2021extreme}.
Parallel computing frameworks utilizing GPU architectures can use Open Computing Language (OpenCL), Compute Unified Device Architecture (CUDA), and Open Accelerators (OpenACC) \cite{navarro2018physically,niemeyer2014recent}.
A detailed comparison of these programming standards can be found in our previous work \cite{xu2017accelerated,xu2023multi}.
Here, we highlight that, with improved data and task management, the OpenACC is promising for thermal LB simulation on GPUs \cite{calore2015accelerating,blair2015accelerating,calore2016performance}.
As demonstrated in the most recent simulation of fluid flows and heat transfer in the side-heated convection cell \cite{xu2023multi}, using OpenACC on a single GPU, the two-dimension (2D) simulation achieved 1.93 billion lattice updates per second (GLUPS) with a grid number of $8193^{2}$, and the three-dimension (3D) simulation achieved 1.04 GLUPS with a grid number of $385^{3}$, which is more than 76\% of the theoretical maximum performance.

In this paper, we aim to extend our previous OpenACC accelerated LB simulation of single-phase thermal convection \cite{xu2017accelerated,xu2023multi} to particle-laden thermal convection.
In contrast to previous works \cite{xiong2012large,ma2023accelerating} who adopted the immersed boundary method to evaluate the fluid-particle interaction force, we adopt the momentum-exchange method due to its simplicity and robustness \cite{ladd1994numerical,chen2013momentum,wen2014galilean}.
To utilize the computing power of multi-node GPU clusters, we adopt a hybrid OpenACC and Message Passing Interface (MPI) approach, in which the OpenACC accelerates computation on a single GPU, and the MPI synchronizes the information between multiple GPUs.
The rest of this paper is organized as follows.
In Section \ref{sec:numericalMethod}, we introduce numerical details of the LB simulation for particle-laden thermal convection.
In Section \ref{sec:singleGPU}, we describe the implementation and optimization details for a single GPU.
In Section \ref{sec:multiGPU}, we describe the implementation and optimization for multi-GPUs.
In Section \ref{sec:conclusions}, the main findings of this work are summarized.

\section{Numerical method \label{sec:numericalMethod}}
\subsection{The LB model for fluid flow and heat transfer}
We simulate thermal convection based on the Boussinesq approximation.
We assume the fluid flow is incompressible, and we treat the temperature as an active scalar that influences the velocity field through the buoyancy.
We neglect viscous heat dissipation and compression work, and assume all the transport coefficients to be constants.
Then, the governing equations can be written as
\begin{subequations}
    \begin{gather}
        \nabla \cdot \mathbf{u}=0 \\
        \frac{\partial \mathbf{u}}{\partial t}+\mathbf{u} \cdot \nabla \mathbf{u}=-\frac{1}{\rho_0} \nabla P+\nu \nabla^2 \mathbf{u}+g \beta_T\left(T-T_0\right) \hat{\mathbf{y}} \\
        \frac{\partial T}{\partial t}+\mathbf{u} \cdot \nabla T=\alpha_T \nabla^2 T
    \end{gather}
    \label{eq:1}
\end{subequations}
where $\mathbf{u}$, $P$ and $T$ are the velocity, pressure, and temperature of the fluid, respectively.
$\rho_0$ and $T_0$ are reference density and temperature, respectively.
$\nu$, $\beta_T$ and $\alpha_T$ denote the viscosity, thermal expansion coefficient, and thermal diffusivity of the fluid, respectively.
$\hat{\mathbf{y}}$ is the unit vector parallel to gravity.
With the scaling
\begin{equation}
    \begin{aligned}
    &\mathbf{x}^*=\mathbf{x} / L_{0}, \quad
    t^*=t / \left( L_{0}^{2}/\alpha_{T} \right), \quad
    \mathbf{u}^*=\mathbf{u} / \left( \alpha_{T}/L_{0} \right), \\
    &P^*=P /\left(\rho_0 \alpha_{T}^{2}/L_{0}^{2}  \right), \quad T^*=\left(T-T_0\right) / \Delta_T
    \end{aligned}
\end{equation}
Then, Eq. (\ref{eq:1}) can be rewritten in dimensionless form as
\begin{subequations}
    \begin{gather}
        \nabla \cdot \mathbf{u}^*=0 \\
        \frac{\partial \mathbf{u}^*}{\partial t^*}+\mathbf{u}^* \cdot \nabla \mathbf{u}^*=-\nabla P^*+Pr \nabla^2 \mathbf{u}^*+Gr Pr^{2}T^* \hat{\mathbf{y}} \\
        \frac{\partial T^*}{\partial t^*}+\mathbf{u}^* \cdot \nabla T^*=\nabla^2 T^*
    \end{gather}
\end{subequations}
Here, $L_{0}$ is the characteristic length and $\Delta_T$ is the temperature difference.
Two independent dimensionless parameters are the Prandtl number ($Pr$) and the Grashof number ($Gr$), which are defined as
\begin{equation}
    Pr=\frac{\nu}{\alpha_T}, \ \ \ Gr =\frac{g \beta_T \Delta_T L_{0}^3}{\nu^{2}}
\end{equation}
Note a third dimensionless parameter of the Rayleigh number ($Ra$) can be calculated as  $Ra=Pr \cdot Gr$.

We adopt the double distribution function (DDF)-based LB model to simulate thermal convective flows with the Boussinesq approximation  \cite{yoshida2010multiple,chai2013lattice,wang2013lattice,contrino2014lattice}.
Specifically, we chose a D2Q9 discrete lattice in two-dimension (2D) for the Navier–Stokes equations to simulate fluid flows, and a D2Q5 discrete lattice in 2D for the energy equation to simulate heat transfer.
To enhance the numerical stability, the multi-relaxation-time (MRT) collision operator is adopted in the evolution equations of both density and temperature distribution functions.
The evolution equation of the density distribution function is written as
\begin{equation}
f_i\left(\mathbf{x}+\mathbf{e}_i \delta_t, t+\delta_t\right)-f_i(\mathbf{x}, t)=-\left(\mathbf{M}^{-1} \mathbf{S}\right)_{i j}\left[\mathbf{m}_j(\mathbf{x}, t)-\mathbf{m}_j^{(\mathrm{eq})}(\mathbf{x}, t)\right]+\delta_t F_i^{\prime} \label{eq:f}
\end{equation}
where $f_i$ is the density distribution function.
$\mathbf{x}$ is the fluid parcel position, $t$ is the time, $\delta_t$ is the time step.
$\mathbf{e}_i$ is the discrete velocity along the $i$th direction.
The macroscopic density $\rho$ and velocity $\mathbf{u}$ are obtained from $\rho=\sum_{i=0}^{q-1} f_i, \mathbf{u}=\left(\sum_{i=0}^{q-1} \mathbf{e}_i f_i+\frac{1}{2} \mathbf{F}\right)/\rho$, where $\mathbf{F}=\rho g \beta_{T}(T-T_{0})\hat{\mathbf{y}}$ is the buoyancy force.

The evolution equation of the temperature distribution function is written as
\begin{equation}
g_i\left(\mathbf{x}+\mathbf{e}_i \delta_t, t+\delta_t\right)-g_i(\mathbf{x}, t)=-\left(\mathbf{N}^{-1} \mathbf{Q}\right)_{i j}\left[\mathbf{n}_j(\mathbf{x}, t)-\mathbf{n}_j^{(\mathrm{eq})}(\mathbf{x}, t)\right] \label{eq:g}
\end{equation}
where $g_i$ is the temperature distribution function.
The macroscopic temperature $T$ is obtained from $T=\sum_{i=0}^{q-1} g_i$.
More numerical details of the thermal LB method can be found in our previous work \cite{xu2017accelerated,xu2023multi,xu2019lattice,xu2021tristable,xu2023wall}.

\subsection{Kinematic model of the solid particle}

We consider the solid particle as a rigid body and its kinematics include translational and rotational motion.
Specifically, we determine the translational motion of the solid particle using Newton's second law as
\begin{equation}
M_{p}\frac{d\mathbf{U}_{c}(t)}{dt}=\mathbf{F}_{p}(t)
\end{equation}
where $M_{p}$  is the mass of the particle, $\mathbf{U}_{c}$ is the velocity of the particle center and  $\mathbf{F}_{p}$ is the total force exerted on the solid particle.
The rotational motion of the solid particle is determined by Euler's second law as
\begin{equation}
\mathbf{I}_{p}\cdot \frac{d\mathbf{\Omega}(t)}{dt}+\mathbf{\Omega}(t)\times[\mathbf{I}_{p}\cdot\mathbf{\Omega}(t)]=\mathbf{T}_{p}(t)
\end{equation}
where $\mathbf{I}_{p}$ is the inertial tensor of the particle, $\mathbf{\Omega}$ is the angular velocity, and $\mathbf{T}_{p}$ is the torque exerted on the solid particle.
In the simulation, we advance the fluid flows and the motion of the particles simultaneously. In other words, the time step used to update fluid and temperature fields is the same as that used for particles' kinematic.

\subsection{Boundary conditions at the fluid-particle interface}

At the particle's curved surface, we adopt the interpolated bounce-back scheme to guarantee no-slip velocity boundary conditions.
A parameter $q=|\mathbf{x}_{f}-\mathbf{x}_{w}|/|\mathbf{x}_{f}-\mathbf{x}_{b}|$ is used to describe the fraction of fluid region in a grid spacing intersected by the solid surface, where $\mathbf{x}_{f}$ is the fluid node near the boundary, $\mathbf{x}_{b}$ is solid node near the boundary, and  $\mathbf{x}_{w}$ is wall interface.
Based on the relative location of $\mathbf{x}_{w}$ between $\mathbf{x}_{f}$  and  $\mathbf{x}_{b}$, we adopt a quadratic interpolation scheme of the density distribution function, which is given as \cite{bouzidi2001momentum}:
for $q \le 0.5$
\begin{equation}
        \begin{split}
 f_{\bar{i}}(\mathbf{x}_{f},t+\delta_{t})= &
q(2q+1)f_{i}^{+}(\mathbf{x}_{f},t)
+(1-4q^{2})f_{i}^{+}(\mathbf{x}_{f}-\mathbf{e}_{i}\delta_{t},t) \\
& -q(1-2q)f_{i}^{+}(\mathbf{x}_{f}-2\mathbf{e}_{i}\delta_{t},t)
+2\omega_{i}\rho_{0}\frac{\mathbf{e}_{\bar{i}}\cdot \mathbf{u}_{w}}{c_{s}^{2}} \\
        \end{split}
\end{equation}
for $q > 0.5$
\begin{equation}
        \begin{split}
f_{\bar{i}}(\mathbf{x}_{f},t+\delta_{t})= &
\frac{1}{q(2q+1)}f_{i}^{+}(\mathbf{x}_{f},t)
+\frac{2q-1}{q}f_{\bar{i}}^{+}(\mathbf{x}_{f},t) \\
& -\frac{2q-1}{2q+1}f_{\bar{i}}^{+}(\mathbf{x}_{f}-\mathbf{e}_{i}\delta_{t},t)
+\frac{1}{q(2q+1)}2\omega_{i}\rho_{0}\frac{\mathbf{e}_{\bar{i}}\cdot \mathbf{u}_{w}}{c_{s}^{2}} \\
        \end{split}
\end{equation}
where $f_{\bar{i}}$ is the distribution function associated with the velocity $\mathbf{e}_{\bar{i}}=-\mathbf{e}_{i}$.
It should be noted that Bouzidi's method requires information from the current fluid nodes as well as its adjacent nodes
(i.e., information at  $\mathbf{x}_{f}$, $\mathbf{x}_{f}-\mathbf{e}_{i}\delta_{t}$, and $\mathbf{x}_{f}-2\mathbf{e}_{i}\delta_{t}$), which poses a challenge to the local computation property of the LB method.
To address this issue, an alternative method is a single-node second-order curved boundary condition \cite{zhao2017single,tao2018one}.
The accuracy, stability, and parallel efficiency of those single-node methods for particle-resolved simulation deserve further comprehensive investigation.

Meanwhile, we assume the fluids and particle temperatures are equal to a constant $T_{w}$ at the surface of the particle.
Then the bounce-back scheme for temperature distribution function   at curved wall boundaries is given as \cite{li2013boundary,li2017lattice}
\begin{equation}
    \begin{split}
g_{\bar{i}}(\mathbf{x}_{f},t+\delta_{t})=
& \left[ c_{d1}g_{i}^{+}(\mathbf{x}_{f},t)+c_{d2}g_{i}^{+}(\mathbf{x}_{f}-\mathbf{e}_{i}\delta_{t},t)+c_{d3}g_{\bar{i}}^{+}(\mathbf{x}_{f},t) \right] \\
& +c_{d4}(2\omega_{i}T_{w}) \\
    \end{split}
\end{equation}
where $g_{\bar{i}}$ is the distribution function associated with the velocity $\mathbf{e}_{\bar{i}}=-\mathbf{e}_{i}$,
and $g_{i}^{+}$ is the post-collision distribution function.
The coefficients $c_{d,1-4}$ are given as
\begin{equation}
c_{d1}=-1, \ \ c_{d2}=\frac{2q-1}{2q+1}, \ \ c_{d3}=\frac{2q-1}{2q+1}, \ \ c_{d4}=\frac{2}{2q+1}
\end{equation}

\subsection{Interaction between fluid and particle phases}

We adopt the momentum-exchange method to calculate the force and torque exerted by the fluid on the solid particle due to its simplicity and robustness \cite{ladd1994numerical}.
Specifically, the hydrodynamic force acting on the solid surface is obtained by summing up the local momentum exchange of the fluid parcels during the bounce-back process.
Because the original momentum-exchange method proposed by Ladd \cite{ladd1994numerical} lacks local Galilean invariance \cite{clausen2009galilean}, we employ a modified momentum-exchange method that simply introduces the relative velocity into the interfacial momentum transfer \cite{wen2014galilean}, then the total hydrodynamic force is calculated as
\begin{equation}
\mathbf{F}=\sum_{\mathbf{x}_{f}} \sum_{i_{bl}} \left[f_{i}^{+}(\mathbf{x}_{f},t)(\mathbf{e}_{i}-\mathbf{u}_{w})-f_{\bar{i}}(\mathbf{x}_{f},t+\delta_{t})(\mathbf{e}_{\bar{i}}-\mathbf{u}_{w})\right]
\label{eq.force}
\end{equation}
and the total torque is calculated as
\begin{equation}
\mathbf{T}=\sum_{\mathbf{x}_{f}} \sum_{i_{bl}}
(\mathbf{x}_{w}-\mathbf{x}_{c})\times \left[f_{i}^{+}(\mathbf{x}_{f},t)(\mathbf{e}_{i}-\mathbf{u}_{w})-f_{\bar{i}}(\mathbf{x}_{f},t+\delta_{t})(\mathbf{e}_{\bar{i}}-\mathbf{u}_{w})\right]
\label{eq.torque}
\end{equation}

A pair of particles will collide when their distance is small.
We adopt the artificial repulsive force model to prevent overlap between the computational particles.
Here, we choose the spring force model that generates a strong repulsive force pushing the two collision particles apart \cite{feng2004immersed}, then the repulsive force is given as
\begin{equation}
\mathbf{F}_{\text{repulsive}}=
    \begin{cases}
    0 & \text{if $|\mathbf{x}_{s}|>s$},\\
    \frac{C}{\varepsilon_{w}}\left(\frac{|\mathbf{x}_{s}|-s}{s} \right)^{2}\frac{\mathbf{x}_{s}}{|\mathbf{x}_{s}|} & \text{if $|\mathbf{x}_{s}|\le s$}.
    \end{cases}
\end{equation}
Here, $\varepsilon_{w}$  represents the stiffness parameter and $s$ represents the threshold distance.
We choose  $\varepsilon_{w}=0.001$ and $s = 3$ l.u., where l.u. denotes the lattice length-unit in the LB simulation  \cite{huang2015multiphase}.
$C=\pi r_{p}^{2}(\rho_{p}-\rho_{f})g$ is the force scale,  $\mathbf{x}_{s}$ denotes the vector with the smallest norm value that points from the wall to the particle.

\subsection{Refilling scheme to construct unknown distribution functions \label{sec:refilling}}
When the particle moves relative to the fixed grids, a solid node may be uncovered by the particle and become a fluid node.
To determine unknown density distribution functions for this new 'born' fluid node, we use the normal extrapolation refilling scheme with velocity constraint \cite{lallemand2003lattice,peng2016implementation}.
Specifically, we first determine the direction of a discrete velocity $\mathbf{e}_{c}$  that maximizes the quantity $\mathbf{n}\cdot\mathbf{e}_{c}$, where $\mathbf{n}$  is the outward normal vector of the wall at the newborn fluid node.
Then the unknown density distribution functions at the newborn fluid node are obtained by a quadratic extrapolation
\begin{align}
f_{\alpha}(\mathbf{x}_{new},t+\delta _{t})=&3f_{\alpha}(\mathbf{x}_{new}+
\mathbf{e}_{c}\delta _{t},t+\delta _{t}) -3f_{\alpha}(\mathbf{x}_{new}+2
\mathbf{e}_{c}\delta _{t},t+\delta _{t})\nonumber \\
&\quad  +f_{\alpha}(\mathbf{x}_{new}+3
\mathbf{e}_{c}\delta _{t},t+\delta _{t})
\end{align}
In the MRT framework, the velocity at the new fluid node can be constrained to the wall velocity via enforcing the momentum moments, which can reduce the fluctuations in the fluid-particle forces \cite{peng2016implementation}.

Meanwhile, to determine unknown temperature distribution functions for the new 'born' fluid node, we use the equilibrium refilling scheme \cite{lallemand2003lattice,rosemann2019comparison} for simplify, which is
\begin{equation}
g_{\alpha}(\mathbf{x}_{new},t+\delta_{t})=g_{\alpha}^{eq}(T_{w},\mathbf{u}_{w})
\end{equation}

\subsection{Validation of the particle-resolved LB model}
The accuracy of the particle-resolved LB model described above has been validated in our previous work \cite{xu2018thermal} via simulation of an elliptical particle settling in isothermal fluids, and a cold circular particle settling in a hot fluid.
Furthermore, our results on an elliptical particle settling in thermal fluids have been confirmed by Walayat \emph{et al.} \cite{walayat2018dynamics} as well as  Suzuki \emph{et al.} \cite{suzuki2020numerical}.
Additionally, we compare the simulation results obtained from the multi-GPU simulation with those of the CPU simulation in Appendix A.
Moreover, we validate the model for particle-particle interaction via simulating the draft-kissing-tumbling (DKT) process with thermal convection in Appendix B.
In the following subsections, we will validate the model by simulating the sedimentation of multiple particles.

\subsubsection{Sedimentation of a group of 800 circular particles in isothermal fluids}

\begin{figure}
  \centering
  \includegraphics[width=0.9\textwidth]{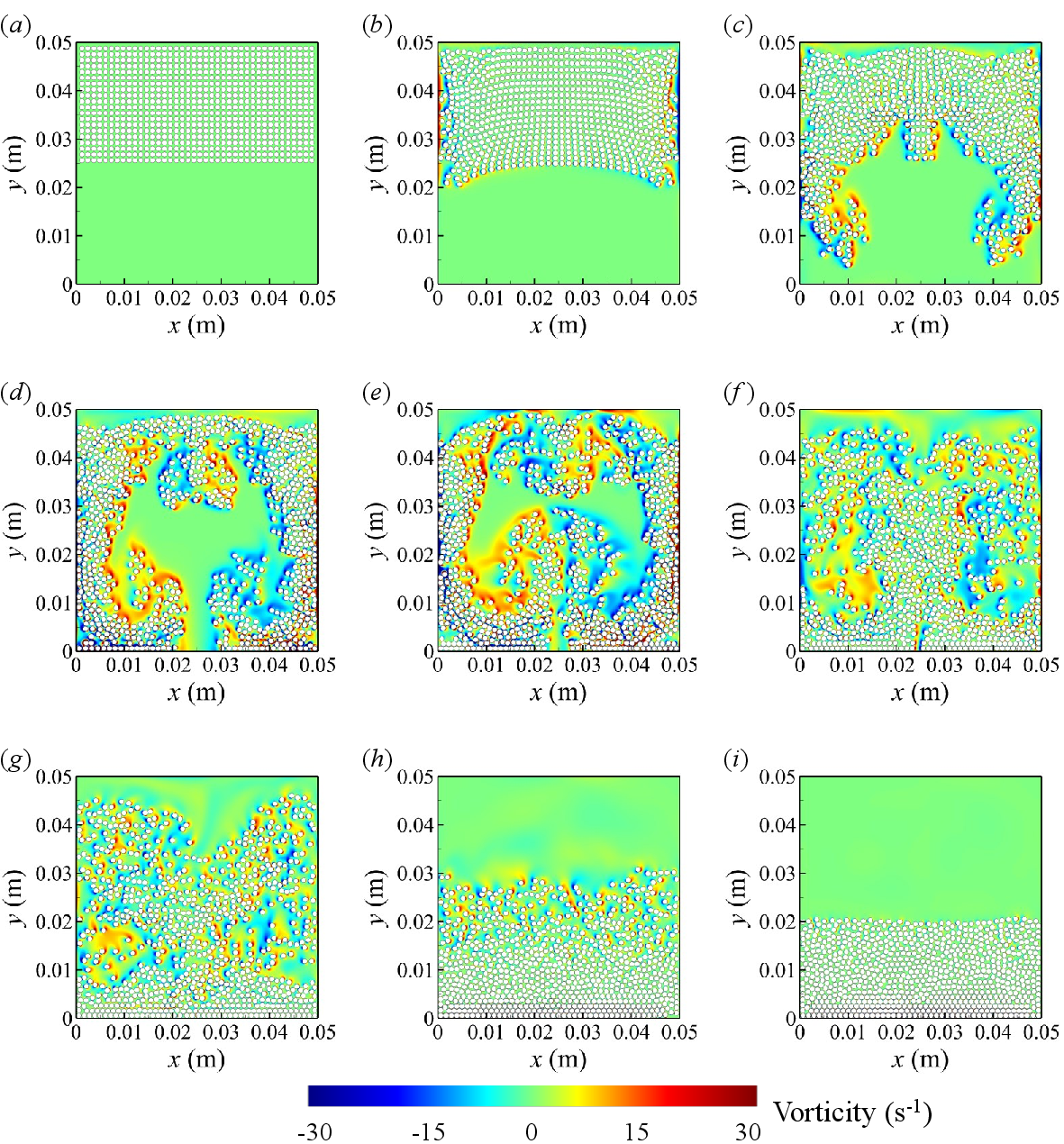}\\
  \caption{Contour of vorticity field during the sedimentation of 800 circular particles in isothermal fluids.
  (\textit{a}) $t = 0$ s, (\textit{b}) $t = 3.125$ s, (\textit{c}) $t = 5$ s, (\textit{d}) $t = 6.25$ s, (\textit{e}) $t = 6.875$ s, (\textit{f}) $t = 8.75$ s, (\textit{g}) $t = 10$ s, (\textit{h}) $t = 17.5$ s, (\textit{i}) $t = 25$ s.}\label{Figure_isothermal_sediments}
\end{figure}

Initially, the particles are placed in the upper part of a square cavity with dimensions of 5 cm $\times$ 5 cm.
All four walls of the cavity are imposed with no-slip velocity boundary conditions.
There are 20 lines of particles, each containing 40 particles, resulting in a total of 800 particles [see Fig. \ref{Figure_isothermal_sediments}(a)].
Each particle has a density of  $\rho_{p}= 1.1$ g/cm$^3$ and a diameter of $d_{p}$ = 1 mm, resulting in a particle volume fraction of 25.1\%.
To mimic the working fluid of water, we set its viscosity as $\nu_{f} = 10^{-6}$ m$^{2}$/s and its density as $\rho_{f}=1$ g/cm$^{3}$.
Following Yu \emph{et al.} \cite{yu2002viscoelastic}, we adopt a reference velocity  $U_{ref}=\sqrt{g\pi d_{p}(\rho_{p}-\rho_{f})/(2\rho_{f})}$ to define the particle Reynolds number $Re_{p}=U_{ref}d_{p}/\nu_{f}$, which is related to the Archimedes number  $Ar=\sqrt{gd_{p}^{3}(\rho_{p}-\rho_{f})/(\nu_{f}^{2}\rho_{f})}$ as  $Re_{p}=\sqrt{\pi/2}Ar$.
In this case, we have $Re_{p} = 12.41$ and $Ar = 9.90$.
A detailed setting for simulation parameters is listed in Table \ref{tb:isothermal_sediments}.
In Fig. \ref{Figure_isothermal_sediments}, we present the contours of the vorticity field during particle sedimentation.
The settling of particle clusters was initially relatively uniform, although particles nearer to the wall exhibited faster settling due to the influence of wall vorticity at $t =3.125$ s [see Fig. \ref{Figure_isothermal_sediments}(b)].
Over time, the particles gradually assumed distinct shapes, with an umbrella-like formation appearing at $t = 5$ s [see Fig. \ref{Figure_isothermal_sediments}(c)], followed by the emergence of a bubble-like shape region in the center of the cavity's lower half at $t = 6.25$ s [see Fig. \ref{Figure_isothermal_sediments}(d)].
As the settling process continued, the particles began to converge in the center of the lower wall [see Fig. \ref{Figure_isothermal_sediments}(e)], and eventually ruptured the bubble region of the cavity [see Figs. \ref{Figure_isothermal_sediments}(f) an \ref{Figure_isothermal_sediments}(g)].
Ultimately, the particles accumulated at the bottom of the cavity, and the fluid gradually returned to stationary state [see Fig. \ref{Figure_isothermal_sediments}(h)].
After 25 s, the particles settled entirely and packed on the bottom wall [see Fig. \ref{Figure_isothermal_sediments}(i)].
The overall patterns observed in the simulations are consistent with previous studies \cite{tao2018combined,feng2004immersed,wan2006direct}.

\begin{table}
\centering
\caption{Simulation parameters for the sedimentation of circular particles in isothermal fluids.
Here, l.u. denotes the lattice length-unit and t.s. denotes the lattice time-step in the LB simulation \cite{huang2015multiphase}.
The length unit conversion is $l_{*} = 2.5\times 10^{-5}$ m/l.u.,
and the time unit conversion is $t_{*} = 3.125\times 10^{-5}$ s/t.s.}
\begin{adjustbox}{center, max width=\textwidth}
\small
\begin{tabular}{cccc}
  \hline
              & Physical system & LB system & Unit conversion \\
  \hline
  Domain size & $W\times H=$ 5 cm $\times$ 5 cm & $\overline{W} \times \overline{H}=$ 2000 l.u. $\times$ 2000 l.u. & $\mathbf{x}=\overline{\mathbf{x}}\cdot l_{*}$ \\
  Particle diameter    & $d_{p}=1$ mm                       & $\overline{d_{p}}=40$ l.u. & $d_{p}=\overline{d_{p}}\cdot l_{*}$ \\
  Kinematic viscosity  & $\nu_{f} = 10^{-6}$ m$^{2}$/s	    & $\overline{\nu_{f}}$ = 0.05 l.u.$^{2}$/t.s. & $\nu_{f}=\overline{\nu_{f}}\cdot l_{*}^{2}/t_{*}$ \\
  Gravity acceleration & $g = 9.8$ m/s$^{2}$	            & $\overline{g}= 3.83 \times 10^{-4}$ l.u./t.s.$^{2}$  & $g=\overline{g}\cdot l_{*}/t_{*}^{2}$ \\
  \hline
\end{tabular} \label{tb:isothermal_sediments}
\end{adjustbox}
\end{table}

\subsubsection{Sedimentation of a group of 172 circular or elliptical hot particles in cold fluids}

\begin{figure}
  \centering
  \includegraphics[width=0.9\textwidth]{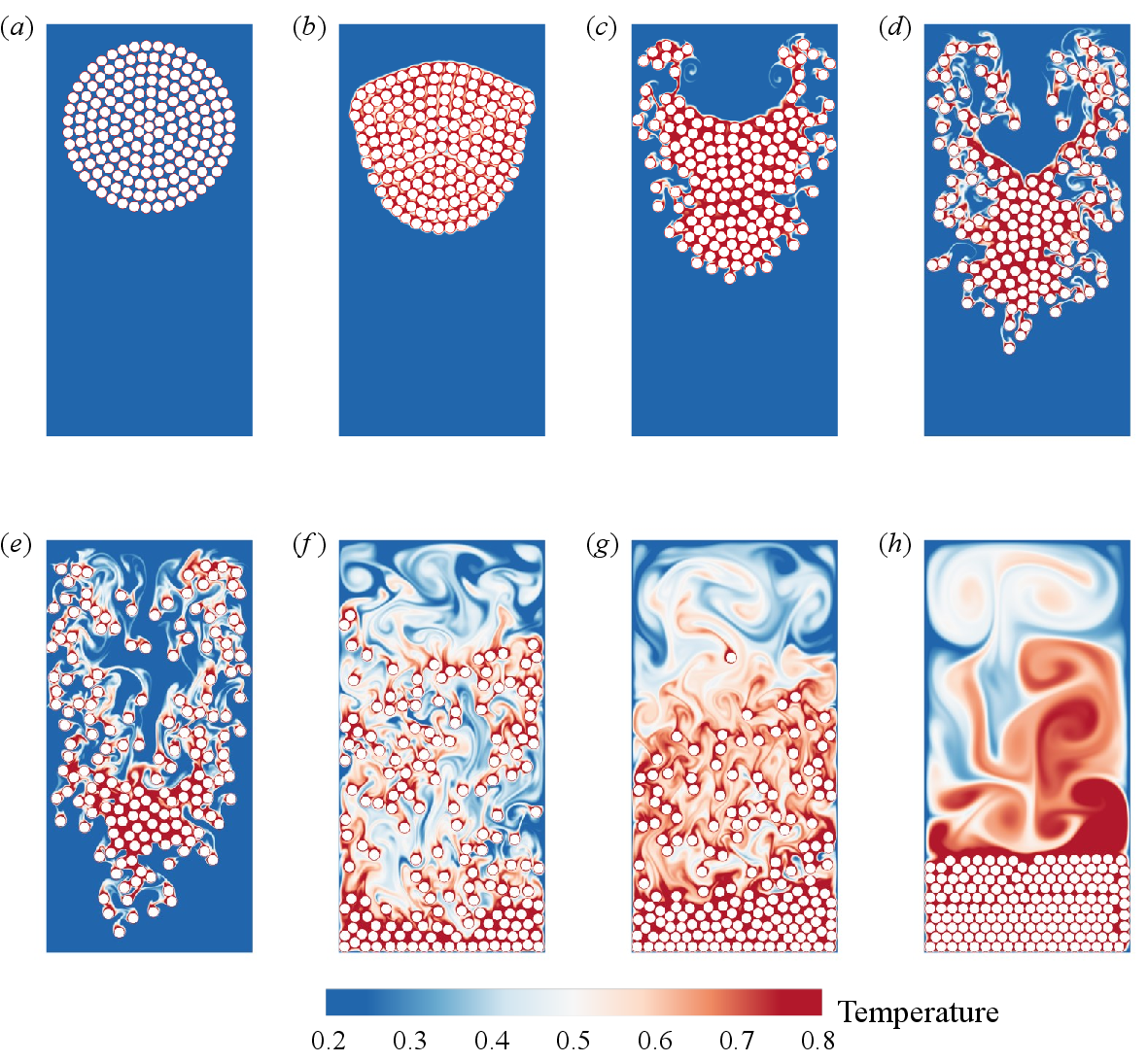}\\
  \caption{Contour of dimensionless temperature field $T^{*}=(T-T_{c})/\Delta_{T}$  during the sedimentation of 172 circular hot particles in cold fluids at the dimensionless time $t^{*}=t/\sqrt{d_{p}/g}$ of
  (\textit{a}) 0, (\textit{b}) 7.42, (\textit{c}) 14.85, (\textit{d}) 22.27, (\textit{e}) 29.70, (\textit{f}) 61.87, (\textit{g}) 84.15, (\textit{h}) 180.67.}\label{Figure_circularThermal}
\end{figure}

We first consider the sedimentation of circular hot particles in cold fluids.
Initially, 172 particles are placed in the upper part of the rectangular cavity, forming a circular cluster.
The size of the cavity is $W \times H$ = 2 cm $\times$ 4 cm, and all four walls of the cavity are imposed no-slip velocity boundary conditions.
Each particle has a density of  $\rho_{p}$ = 1.5 g/cm$^{3}$ and a diameter of $d_{p}$ = 1 mm, resulting in a particle volume fraction of 16.9\%.
The position of the particle cluster is initialized as follows: the center of the cluster is occupied by one particle, located at $(0.5W, 0.75H)$, and seven layers of particles are then spread outward in a uniform ring shape, with the number of particles in each layer being 6, 12, 18, 24, 31, 37, 43 [see Fig. \ref{Figure_circularThermal}(a)].
To mimic the working fluid of water, we set its viscosity as $\nu_{f} = 10^{-6}$ m$^{2}$/s and its density as  $\rho_{f}=1$ g/cm$^{3}$.
We adopt a reference velocity $U_{ref}=\sqrt{g\pi d_{p}(\rho_{p}-\rho_{f})/(2\rho_{f})}$  to define the particle Reynolds number  $Re_{p}=U_{ref}d_{p}/\nu_{f}$, and the Archimedes number is  $Ar=\sqrt{gd_{p}^{3}(\rho_{p}-\rho_{f})/(\nu_{f}^{2}\rho_{f})}$.
In this case, we have $Re_{p} = 87.7$ and $Ar = 70.0$.
We set the surface temperature of the particles to be a constant high-temperature $T = T_{h}$, the wall of the square cavity to be a constant low-temperature $T = T_{c}$, and the initial temperature of the fluid also be a low-temperature $T_{c}$;
thus, the temperature difference is  $\Delta_{T}=T_{h}-T_{c}$.
We choose the particle Grashof number $Gr_{p} = g\beta_{T}\Delta_{T}d_{p}^{3}/\nu_{f}^{2} = 100$, and the Prandtl number $Pr = 5$.
A detailed setting for simulation parameters is listed in Table \ref{tb:circularThermal}.
In Fig. \ref{Figure_circularThermal}, we present the contours of the temperature field during particle sedimentation.
As the particle cluster settles, the particles closest to the vertical centerline experience a faster rate of settling than those located near the walls.
The walls impede the settling of the particles, causing the cluster to contort into a pine cone shape [see Fig. \ref{Figure_circularThermal}(b)].
Subsequently, the particles close to the walls are swept upward, elongating the cluster into a crescent shape [see Figs. \ref{Figure_circularThermal}(c) and \ref{Figure_circularThermal}(d)].
As the sedimentation continues, the particle cluster becomes an irregular shape with the generation of numerous thermal plumes [see Figs. \ref{Figure_circularThermal}(e) and \ref{Figure_circularThermal}(f)], and these plumes play a significant role in pushing and mixing the particles.
Eventually, most of the particles settle at the bottom of the cavity; however, due to buoyancy, some particles still float for an extended period before they can finally settling [see Figs. \ref{Figure_circularThermal}(g) and \ref{Figure_circularThermal}(h)].
The overall patterns observed in the simulations are consistent with previous studies \cite{walayat2018efficient,feng2008inclusion,tao2022sharp}.

\begin{table}
\centering
\caption{Simulation parameters for the sedimentation of circular hot particles in cold fluids.
Here, t.u. denotes the temperature unit in the LB simulation \cite{huang2015multiphase}.
The length unit conversion is $l_{*} = 2.5\times 10^{-5}$ m/l.u.,
the time unit conversion is $t_{*} = 6.25\times 10^{-6}$ s/t.s.,
and the temperature unit conversion is $T_{*} = 48.6$ K/t.u.}
\begin{adjustbox}{center, max width=\textwidth}
\small
\begin{tabular}{cccc}
  \hline
              & Physical system & LB system & Unit conversion \\
  \hline
  Domain size & $W\times H=$ 2 cm $\times$ 4 cm & $\overline{W} \times \overline{H}=$ 800 l.u. $\times$ 1600 l.u. & $\mathbf{x}=\overline{\mathbf{x}}\cdot l_{*}$ \\
  Particle diameter    & $d_{p}=1$ mm                       & $\overline{d_{p}}=40$ l.u. & $d_{p}=\overline{d_{p}}\cdot l_{*}$ \\
  Kinematic viscosity  & $\nu_{f} = 10^{-6}$ m$^{2}$/s	    & $\overline{\nu_{f}}$ = 0.01 l.u.$^{2}$/t.s. & $\nu_{f}=\overline{\nu_{f}}\cdot l_{*}^{2}/t_{*}$ \\
  Thermal diffusivity  & $\alpha_{T}= 2\times 10^{-5}$ m$^{2}$/s    & $\overline{\alpha_{T}}$ = 0.002 l.u.$^{2}$/t.s. & $\alpha_{T}=\overline{\alpha_{T}}\cdot l_{*}^{2}/t_{*}$ \\
  Gravity acceleration & $g = 9.8$ m/s$^{2}$	            & $\overline{g}= 1.53 \times 10^{-5}$ l.u./t.s.$^{2}$  & $g=\overline{g}\cdot l_{*}/t_{*}^{2}$ \\
  Thermal expansion coefficient	& 	 $ \beta = 2.1\times 10^{-4}$ K$^{-1}$  &  $\overline{\beta} = 1.02\times 10^{-2}$ t.u.$^{-1}$ & $\beta=\overline{\beta}/T_{*}$ \\
  Temperature difference	& 	 $\Delta_{T} = 48.6$ K  &  $\overline{\Delta_{T}}$ = 1 t.u. & $\Delta_{T}=\overline{\Delta_{T}}\cdot T_{*}$ \\
  \hline
\end{tabular} \label{tb:circularThermal}
\end{adjustbox}
\end{table}

\begin{figure}
  \centering
  \includegraphics[width=0.9\textwidth]{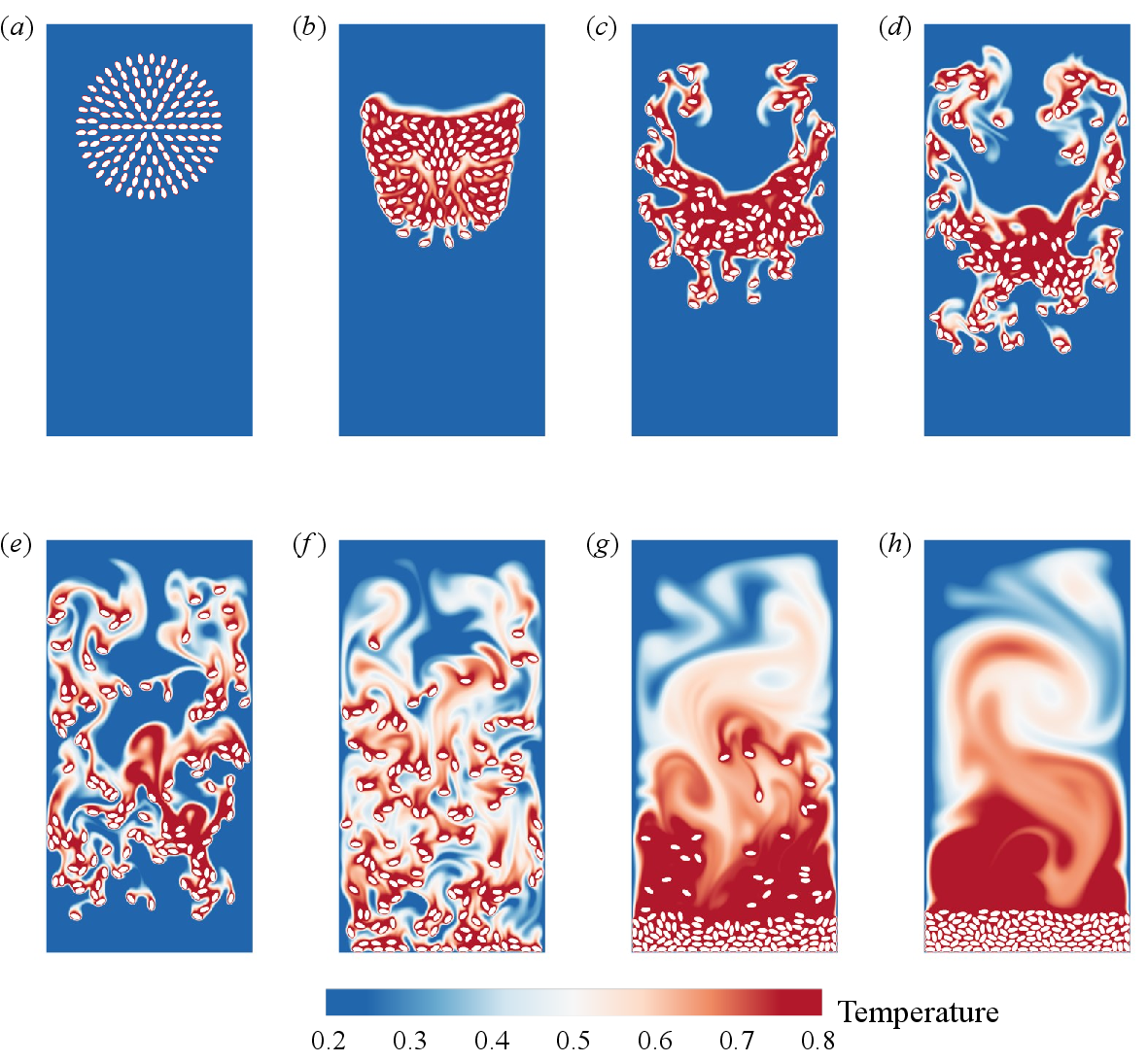}\\
  \caption{Contour of dimensionless temperature field $T^{*}=(T-T_{c})/\Delta_{T}$ during the sedimentation of 129 elliptical hot particles in cold fluids at the dimensionless time $t^{*}=t/\sqrt{A/g}$  of
  (\textit{a}) 0, (\textit{b}) 14.0, (\textit{c}) 28.0, (\textit{d}) 42.0, (\textit{e}) 56.0, (\textit{f}) 84.0, (\textit{g}) 140.0, (\textit{h}) 196.0.}\label{Figure_ellipticalThermal}
\end{figure}

We then consider the sedimentation of elliptical hot particles in cold fluids.
Initially, 129 particles are placed in the upper part of the rectangular cavity, forming a circular cluster.
The size of the cavity is $W \times H$ = 1 cm $\times$ 2 cm, and all four walls of the cavity impose no-slip velocity boundary conditions.
Each particle has a density of  $\rho_{p}$ = 1.5 g/cm$^{3}$, a major axis of $A$ = 0.5 mm and a minor axis of $B$ = 0.25 mm, resulting in a particle volume fraction of 6.3\%.
The position of the particle cluster is initialized as follows: the center of the cluster is occupied by one particle, located at $(0.5W,0.75H)$, and six layers of particles are then spread outward in a uniform ring shape with the number of particles in each layer being 6, 12, 18, 24, 31, 37 [see Fig. \ref{Figure_ellipticalThermal}(a)].
To mimic the working fluid of water, we set its viscosity as $\nu_{f} = 10^{-6}$ m$^{2}$/s and its density as $\rho_{f}=1$ g/cm$^{3}$.
We adopt a reference velocity $U_{ref}=\sqrt{g\pi A(\rho_{p}-\rho_{f})/(2\rho_{f})}$  to define the particle Reynolds number  $Re_{p}=U_{ref}d_{p}/\nu_{f}$, and the Archimedes number is  $Ar=\sqrt{gd_{p}^{3}(\rho_{p}-\rho_{f})/(\nu_{f}^{2}\rho_{f})}$.
In this case, we have $Re_{p} = 31.0$ and $Ar = 24.7$.
We set the surface temperature of the particles to be a constant high-temperature $T = T_{h}$, the wall of the square cavity to be a constant low-temperature $T = T_{c}$, and the initial temperature of the fluid also be a low-temperature $T_{c}$;
thus, the temperature difference is  $\Delta_{T}=T_{h}-T_{c}$.
We choose the particle Grashof number $Gr_{p} = g\beta_{T}\Delta_{T}A^{3}/\nu_{f}^{2} = 10$, and the Prandtl number $Pr = 5$.
A detailed setting for simulation parameters is listed in Table \ref{tb:ellipticalThermal}.
In Fig. \ref{Figure_ellipticalThermal}, we present the contours of the temperature field during particle sedimentation.
We can see that the collective motion of elliptical particles during sedimentation is qualitatively the same as that of circular particles:
the initially placed cluster is contorted into a pine cone shape, followed by elongating into a crescent shape, and then an irregular shape.
Previously, we identified an anomalous rolling mode and an inclined mode for a single hot elliptical particle settling in cold fluids \cite{xu2018thermal}, suggesting particle shape plays a critical role in affecting the sedimentation;
whether there is a new physical mechanism on multiple elliptical particles sedimentation deserves further study within a wide range of control parameters.

\begin{table}
\centering
\caption{Simulation parameters for the sedimentation of elliptical hot particles in cold fluids.
The length unit conversion is $l_{*} = 10^{-5}$ m/l.u.,
the time unit conversion is $t_{*} = 10^{-6}$ s/t.s.,
and the temperature unit conversion is $T_{*} = 38.9$ K/t.u.}
\begin{adjustbox}{center, max width=\textwidth}
\small
\begin{tabular}{cccc}
  \hline
              & Physical system & LB system & Unit conversion \\
  \hline
  Domain size & $W\times H=$ 1 cm $\times$ 2 cm & $\overline{W} \times \overline{H}=$ 1000 l.u. $\times$ 2000 l.u. & $\mathbf{x}=\overline{\mathbf{x}}\cdot l_{*}$ \\
  Particle size    & $A \times B=$ 0.5 mm $\times$ 0.25 mm      & $\overline{A}\times \overline{B}=50$ l.u. $\times$ 25 l.u. & $d_{p}=\overline{d_{p}}\cdot l_{*}$ \\
  Kinematic viscosity  & $\nu_{f} = 10^{-6}$ m$^{2}$/s	    & $\overline{\nu_{f}}$ = 0.01 l.u.$^{2}$/t.s. & $\nu_{f}=\overline{\nu_{f}}\cdot l_{*}^{2}/t_{*}$ \\
  Thermal diffusivity  & $\alpha_{T}= 2\times 10^{-5}$ m$^{2}$/s    & $\overline{\alpha_{T}}$ = 0.002 l.u.$^{2}$/t.s. & $\alpha_{T}=\overline{\alpha_{T}}\cdot l_{*}^{2}/t_{*}$ \\
  Gravity acceleration & $g = 9.8$ m/s$^{2}$	            & $\overline{g}= 9.8 \times 10^{-7}$ l.u./t.s.$^{2}$  & $g=\overline{g}\cdot l_{*}/t_{*}^{2}$ \\
  Thermal expansion coefficient	& 	 $ \beta = 2.1\times 10^{-4}$ K$^{-1}$  &  $\overline{\beta} = 8.16\times 10^{-3}$ t.u.$^{-1}$ & $\beta=\overline{\beta}/T_{*}$ \\
  Temperature difference	& 	 $\Delta_{T} = 38.9$ K  &  $\overline{\Delta_{T}}$ = 1 t.u. & $\Delta_{T}=\overline{\Delta_{T}}\cdot T_{*}$ \\
  \hline
\end{tabular} \label{tb:ellipticalThermal}
\end{adjustbox}
\end{table}

\subsubsection{Sedimentation of 4,800 circular hot particles in cold fluids \label{sec:4800circularHotParticles}}

\begin{figure}
  \centering
  \includegraphics[width=0.9\textwidth]{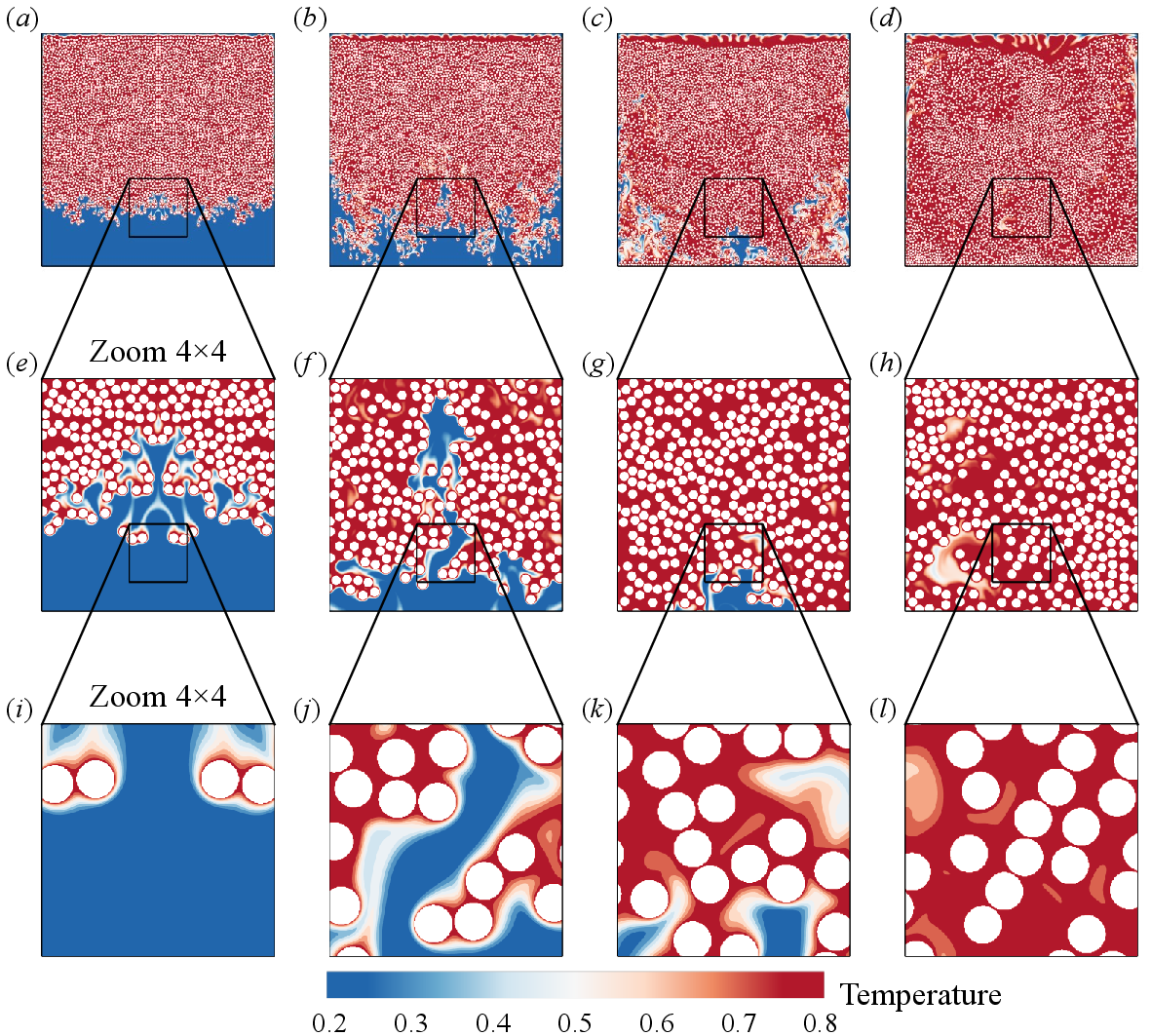}\\
  \caption{Contour of temperature field during the sedimentation of 4,800 circular hot particles in cold fluids at the dimensionless time $t^{*}=t/\sqrt{d_{p}/g}$  of
  (\textit{a}) 70.8350, (\textit{b}) 141.6700, (\textit{c}) 212.5051, (\textit{d}) 283.3401.
  The second row (\textit{e}-\textit{h}) and third row (\textit{i}-\textit{l}) show successive zooms of the area indicated in the black box.}\label{Figure_4800circularHotParticles}
\end{figure}

To better utilize the computing power of the GPUs, we also simulate the sedimentation of a large number of 4,800 circular hot particles in cold fluids.
The size of the cavity is $W \times H$ = 8 cm $\times$ 8 cm, and all four walls of the cavity impose no-slip velocity boundary conditions.
Each particle has a density of  $\rho_{p}$ = 1.1 g/cm$^{3}$ and a diameter of $d_{p}$ = 0.8 mm, resulting in a particle volume fraction of 37.7\%.
Initially, the particles are placed in the upper part of a square cavity, and there are 60 lines of particles with each line having 80 particles [see Fig. \ref{Figure_4800circularHotParticles}(a)].
To mimic the working fluid of water, we set its viscosity as $\nu_{f} = 10^{-6}$ m$^{2}$/s and its density as  $\rho_{f}=$ 1 g/cm$^{3}$.
In this case, we have $Re_{p}=U_{ref}d_{p}/\nu_{f} = 28.1$ and  $Ar=\sqrt{gd_{p}^{3}(\rho_{p}-\rho_{f})/(\nu_{f}^{2}\rho_{f})}= 22.4$.
We set the surface temperature of the particles to be a constant high-temperature $T = T_{h}$, the wall of the square cavity to be a constant low-temperature $T = T_{c}$, and the initial temperature of the fluid also be a low-temperature $T_{c}$;
thus, the temperature difference is  $\Delta_{T}=T_{h}-T_{c}$.
We choose the particle Grashof number $Gr_{p} = g\beta_{T}\Delta_{T}d_{p}^{3}/\nu_{f}^{2} = 60$, and the Prandtl number $Pr = 5$.
A detailed setting for simulation parameters is listed in Table \ref{tb:4800circularHotParticles}.
In Fig. \ref{Figure_4800circularHotParticles}, we present the contours of the temperature field during particle sedimentation.
The particles in a fluid mixture tend to settle due to gravity, and they transfer heat to the surrounding fluid through convection, resulting in the formation of complex flow patterns.
The Rayleigh-Taylor instability is a key mechanism that underlies this process, leading to the formation of finger-like patterns, bifurcation in flow patterns, coalescence, breaking of symmetry, and the development of the draft-kissing-tumbling phenomenon.
This instability arises from the fact that the heavier particles tend to sink faster than the lighter fluid above them, thus, the interface between the two regions becomes unstable and complex vortices are formed.
These vortices, which can vary in size and shape depending on the properties of the fluid mixture and the geometry of the container, play a key role in accelerating the mixing process.
By pushing the particles back up to the top of the cavity, the particles are redistributed and the mixing efficiency is enhanced.
However, as the particles become more evenly distributed, their settling velocity decreases, eventually leading to a stage of slow settling at the bottom of the container.
In Section \ref{sec:singleGPU} and Section \ref{sec:multiGPU}, we will use this problem as the benchmark to measure the parallel performance of the particle-resolved thermal LB simulation.

\begin{table}
\centering
\caption{Simulation parameters for the sedimentation of circular hot particles in cold fluids.
The length unit conversion is $l_{*} = 2\times 10^{-5}$ m/l.u.,
the time unit conversion is $t_{*} = 4\times 10^{-6}$ s/t.s.,
and the temperature unit conversion is $T_{*} = 56.9$ K/t.u.}
\begin{adjustbox}{center, max width=\textwidth}
\small
\begin{tabular}{cccc}
  \hline
              & Physical system & LB system & Unit conversion \\
  \hline
  Domain size & $W\times H=$ 8 cm $\times$ 8 cm & $\overline{W} \times \overline{H}=$ 4000 l.u. $\times$ 4000 l.u. & $\mathbf{x}=\overline{\mathbf{x}}\cdot l_{*}$ \\
  Particle diameter    & $d_{p}=0.8$ mm                       & $\overline{d_{p}}=40$ l.u. & $d_{p}=\overline{d_{p}}\cdot l_{*}$ \\
  Kinematic viscosity  & $\nu_{f} = 10^{-6}$ m$^{2}$/s	    & $\overline{\nu_{f}}$ = 0.01 l.u.$^{2}$/t.s. & $\nu_{f}=\overline{\nu_{f}}\cdot l_{*}^{2}/t_{*}$ \\
  Thermal diffusivity  & $\alpha_{T}= 2\times 10^{-5}$ m$^{2}$/s    & $\overline{\alpha_{T}}$ = 0.002 l.u.$^{2}$/t.s. & $\alpha_{T}=\overline{\alpha_{T}}\cdot l_{*}^{2}/t_{*}$ \\
  Gravity acceleration & $g = 9.8$ m/s$^{2}$	            & $\overline{g}= 7.84\times 10^{-6}$ l.u./t.s.$^{2}$  & $g=\overline{g}\cdot l_{*}/t_{*}^{2}$ \\
  Thermal expansion coefficient	& 	 $ \beta = 2.1\times 10^{-4}$ K$^{-1}$  &  $\overline{\beta} = 1.20\times 10^{-2}$ t.u.$^{-1}$ & $\beta=\overline{\beta}/T_{*}$ \\
  Temperature difference	& 	 $\Delta_{T} = 56.9$ K  &  $\overline{\Delta_{T}}$ = 1 t.u. & $\Delta_{T}=\overline{\Delta_{T}}\cdot T_{*}$ \\
  \hline
\end{tabular} \label{tb:4800circularHotParticles}
\end{adjustbox}
\end{table}

\section{Implementation and optimization on a single GPU \label{sec:singleGPU}}

A naïve parallel approach for calculating fluid-particle interactions (i.e., implementing Eqs. (\ref{eq.force}) and (\ref{eq.torque}) to obtain hydrodynamic force and torque, respectively) is to use one GPU thread per computational grid, with 128 threads working together to handle the calculation.
However, this approach can lead to load imbalance issues because only fluid nodes near the particle surface perform the calculation.
When 128 grids are processed simultaneously, the threads that are not assigned tasks to calculate fluid-particle interaction have to wait for all other threads to complete their calculations.
To address this load imbalance issue, we adopt the indirect addressing method inspired by simulating flows in porous media \cite{pan2004high,huang2015multi}.
Specifically, we collect the grid lines involved in calculating fluid-particle interaction into a contiguous array in memory, including the position indices (i.e., $i$ and $j$ in the two-dimensional domain) of fluid nodes involved in the hydrodynamic force and torque computation, as well as the discrete velocity direction (i.e., $\alpha$) across the fluid-particle boundary.
The advantage of this method is that only indexes of the participating computational variables are stored, and these indexed flow variables are all involved in the computation to avoid the load imbalance issue.
In particle-laden flow, due to the motion of particles, the grid link connections may change at each timestep, and the index information must be collected simultaneously.
This is different from simulating flow in porous media, where the solid skeleton is stationary, and the indirectly addressed indexes only need to be calculated once throughout the simulation.

\begin{figure}
  \centering
  \includegraphics[width=0.6\textwidth]{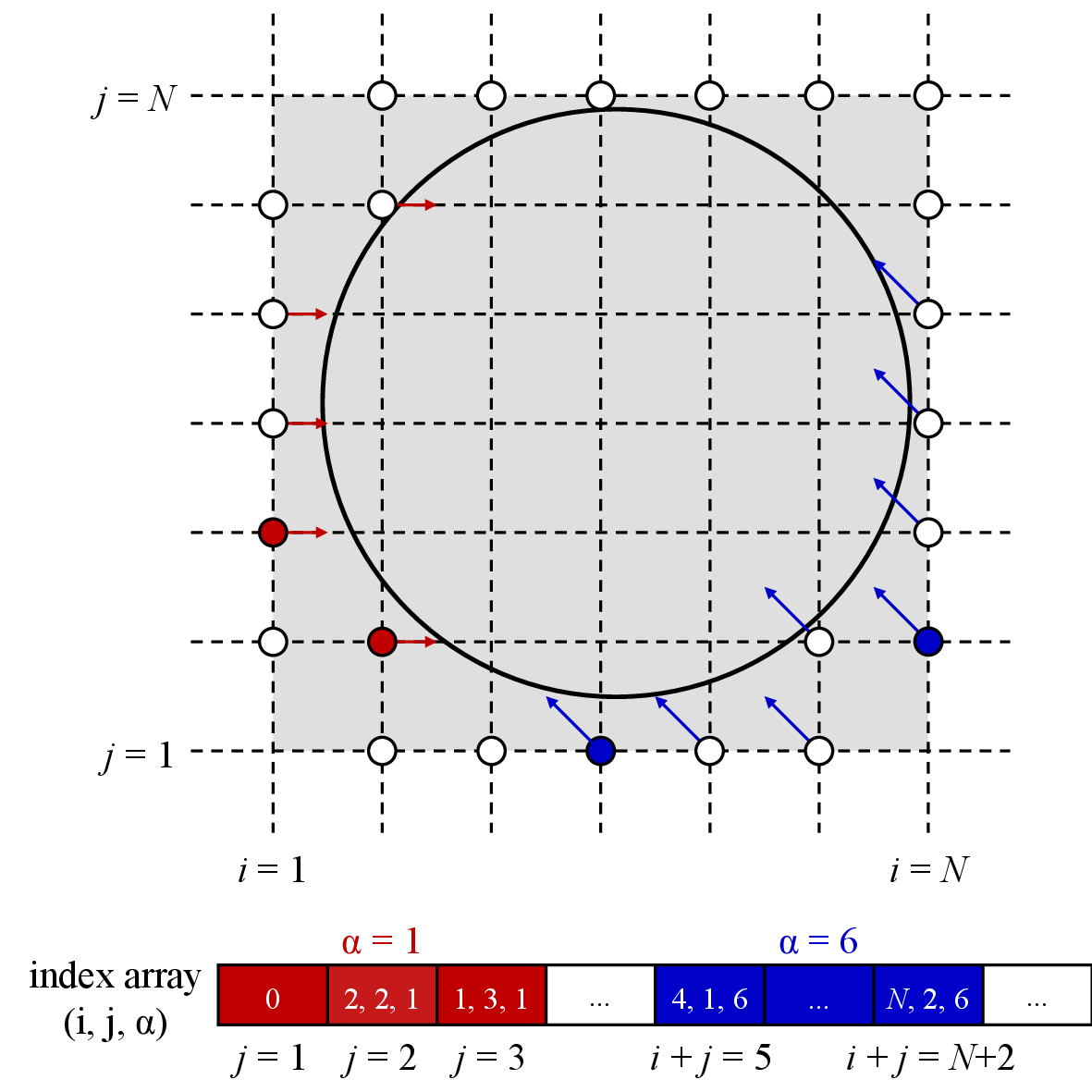}\\
  \caption{Fixed mapping from solid-fluid grid linkage to continuous memory arrays.
  The gray area represents the effective zone with a size of $N \times N$, where $N = INT(x_{c}+d_{p}/2)- INT(x_{c}-d_{p}/2)+3$.}\label{Figure_indirectAddressing}
\end{figure}

In addition, we establish a fixed mapping between the fluid-particle grid and the continuous memory array.
As illustrated in Fig. \ref{Figure_indirectAddressing}, for a particle whose center is $(x_{c}, y_{c})$, we only consider whether a fluid node possesses boundary links for $x$-coordinate ranges from $INT(x_{c}-d_{p}/2)-1$ to $INT(x_{c}+d_{p}/2)+2$ and $y$-coordinate ranges from $INT(y_{c}-d_{p}/2)-1$ to $INT(y_{c}+d_{p}/2)+2$.
Here, $INT(d_{p})$ is the largest integer not exceeding the magnitude of $d_{p}$.
Previous results shown that this boundary link search algorithm is significantly faster than blindly comparing each node to each particle directly \cite{gao2013lattice}.
For a discrete velocity parallel to the grid line (e.g., $\mathbf{e}_{1}$), there will be $N$ lines crossing the effective domain, with vertical coordinates $j = 1, \cdots, N$.
Here, $N = INT(x_{c}+d_{p}/2)- INT(x_{c}-d_{p}/2)+3$ denotes the length of the effective zone.
Each line will have at most one fluid-particle interaction node pointing in the direction of $\mathbf{e}_{1}$.
For a discrete velocity diagonal to the grid line (e.g., $\mathbf{e}_{6}$), there will be $2N - 3$ lines crossing the effective domain and their coordinates meet the relation of $i + j = 3, 4, \cdots, 2N-1$.
Similarly, each line will have at most one fluid-particle interaction node pointing in the direction of $\mathbf{e}_{6}$.
Thus, based on the D2Q9 discrete velocity model, the maximum number of indexes collected across the entire influence domain is $4\times N+4\times (2N-3)=12N-12$.
Although a few of these indexes may be unavailable (e.g., the line with $j = 1$ does not intersect with the particle, resulting in two invalid index positions), it is unlikely to cause severe load imbalance issues.

In Table \ref{tb:direct-indirect}, we compare the running time between direct and indirect addressing methods using the NVIDIA Nsight Systems tool.
Here, we simulate the sedimentation of 4,800 hot particles in cold fluids (see detailed settings in Section \ref{sec:4800circularHotParticles}), and the iteration steps are fixed as 6000.
The analysis involved summing the time consumed by all fluid-related computation steps and particle-related computation steps.
Note subroutines that account for less than 1\% of the total time are neglected.
The results indicate that, although the indirect addressing method requires an additional step for collecting index information, it reduces the time consumption for particle-related computation by a factor of 3.3X compared to the direct addressing method.
This leads to an overall improvement in code performance by 1.4X.
We also adopt the Million Lattice Update Per Second (MLUPS) as the metric to characterize the parallel performance of the LB simulation, which is defined as  \cite{bailey2009accelerating}
\begin{equation}
\text{MLUPS}=\frac{\text{mesh size}\times \text{iteration steps}}{\text{running time}\times 10^{6}}
\end{equation}
We obtain the overall parallel performance of 1209 MLUPS and 1750 MLUPS for the direct and indirect addressing methods, respectively.

\begin{table}
\caption{Comparison of running time between direct and indirect addressing methods. The iteration steps are fixed as 6000.
Data included in the brackets represent the percentage of time consumption.
Note routines that account for less than 1\% of the total time are neglected.}
\centering
\begin{tabular}{cccc}
  \hline
                      & Fluid-related & Particle-related & Overall  \\
  \hline
  Direct addressing   & 41.9 s (52.8\%)       & 37.1 s (46.7\%)          & 79.4 s  \\ 
  Indirect addressing & 43.1 s (78.6\%)       & 11.3 s (20.6\%)          & 54.9 s  \\   
  Speed-up            & 1X                    & 3.3X                     & 1.4X    \\
  \hline
\end{tabular} \label{tb:direct-indirect}
\end{table}

\section{Implementation and optimization on multi-GPUs \label{sec:multiGPU}}

A major limitation in GPU computing is the available device memory; for example, the state-of-art NVIDIA A100 GPU accelerator offers a maximum of 40GB.
A solution to the memory limitation is to use multiple GPUs, where the GPUs are distributed across multiple CPU nodes, and MPI is used to coordinate the computational tasks.
In this work, we conducted experiments on a GPU cluster where each node is equipped with four NVIDIA A100 GPUs.
The network interconnects use 100 Gigabits per second (Gbps) Remote direct memory access over Converged Ethernet (RoCE).
The inter-GPU-GPU communication within a node goes over the PCI-e.
To automatically utilize the GPUDirect acceleration technologies, we adopt a CUDA-aware MPI implemented in OpenMPI.
With GPUDirect technology, including Peer to Peer (P2P) and Remote Direct Memory Access (RDMA), the buffers can be directly sent from a GPU's memory to another GPU's memory or the network without touching the host memory  \cite{ye2022accelerating}.

\subsection{A naïve implementation using hybrid OpenACC and MPI technique}

We adopt a mono-dimensional partitioning of the computational domain and decompose the domain along the $y$-direction.
To facilitate communication between sub-domains, we add three ghost layers outside the boundary of each subdomain to exchange data with adjacent subdomains.
In the particle-related calculation, it is essential to exchange the information of $f^{+}$ and $g^{+}$ at the boundary nodes after the collision step.
When a particle is close to the boundary of a subdomain, the refilling calculation on the new 'born' fluid node requires information from adjunct subdomains to interpolate the distribution function.
After the collision step, each subdomain exchanges the $f^{+}_{1-8}$ at two layers of boundary nodes, and $g^{+}_{1-4}$ at the boundary nodes with their neighbors.
After calculating the fluid-particle interaction, the $f_{0-8}$ at three layers of boundary nodes is exchanged with their neighbors.
Although the computational domain is decomposed into slices and each GPU only accesses the fluid nodes of its assigned subdomain, the information of all particles is shared among all GPUs.
This means that each particle-related loop requires iteration among all particles, leading to a loss of computational efficiency.
The fluid nodes near the particle surface exchange momentum with the particles, and we use the \emph{MPI ALLREDUCE} to obtain the combined forces on the particle in all sub-domains.

\begin{figure}
  \centering
  \includegraphics[width=\textwidth]{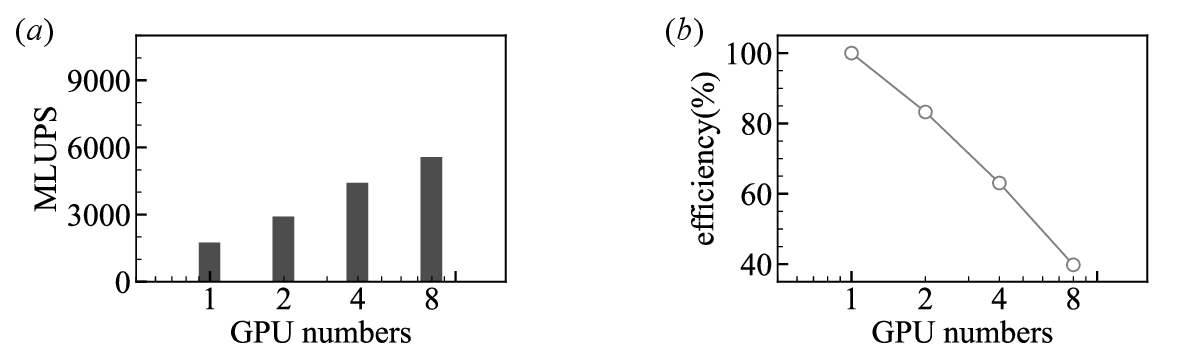}\\
  \caption{Parallel performance in terms of (\textit{a}) the MLUPS and (\textit{b}) the parallel efficiency for a naïve implementation using hybrid OpenACC and MPI technique.}\label{Fig_mono}
\end{figure}

Fig. \ref{Fig_mono} shows the parallel performance in terms of the MLUPS and parallel efficiency.
Parallel efficiency is defined as  $n=T_{1}/(T_{n}\cdot n)$.
Here, $T_{1}$ denotes the running time using a single GPU, and $T_{n}$ denotes the running time using $n$ GPUs.
We can see that with an increase in the number of GPUs, the MULPS generally increases.
Using 8 GPUs, the simulation can achieve 5572 MLUPS, indicating a higher computational capacity.
However, the parallel efficiency degrades when more GPUs are used.
Notably, the parallel efficiency is only 39.8\% using 8 GPUs, indicating the parallel code is not scalable.
Two reasons may be responsible for the poor parallel performance.
First, the presence of redundant data exchange, such as the exchange of distribution functions on three layers of ghost nodes, can increase communication overhead.
Secondly, although the computational domain has been decomposed into slices and each GPU is only responsible for updating fluid node information within its subdomain, information on the particle group is required on every GPU, leading to additional communication overhead.
To further boost the parallel performance using multi-GPUs, we describe some optimization strategies in the following subsections.

\subsection{Building the domain list}

\begin{figure}
  \centering
  \includegraphics[width=0.6\textwidth]{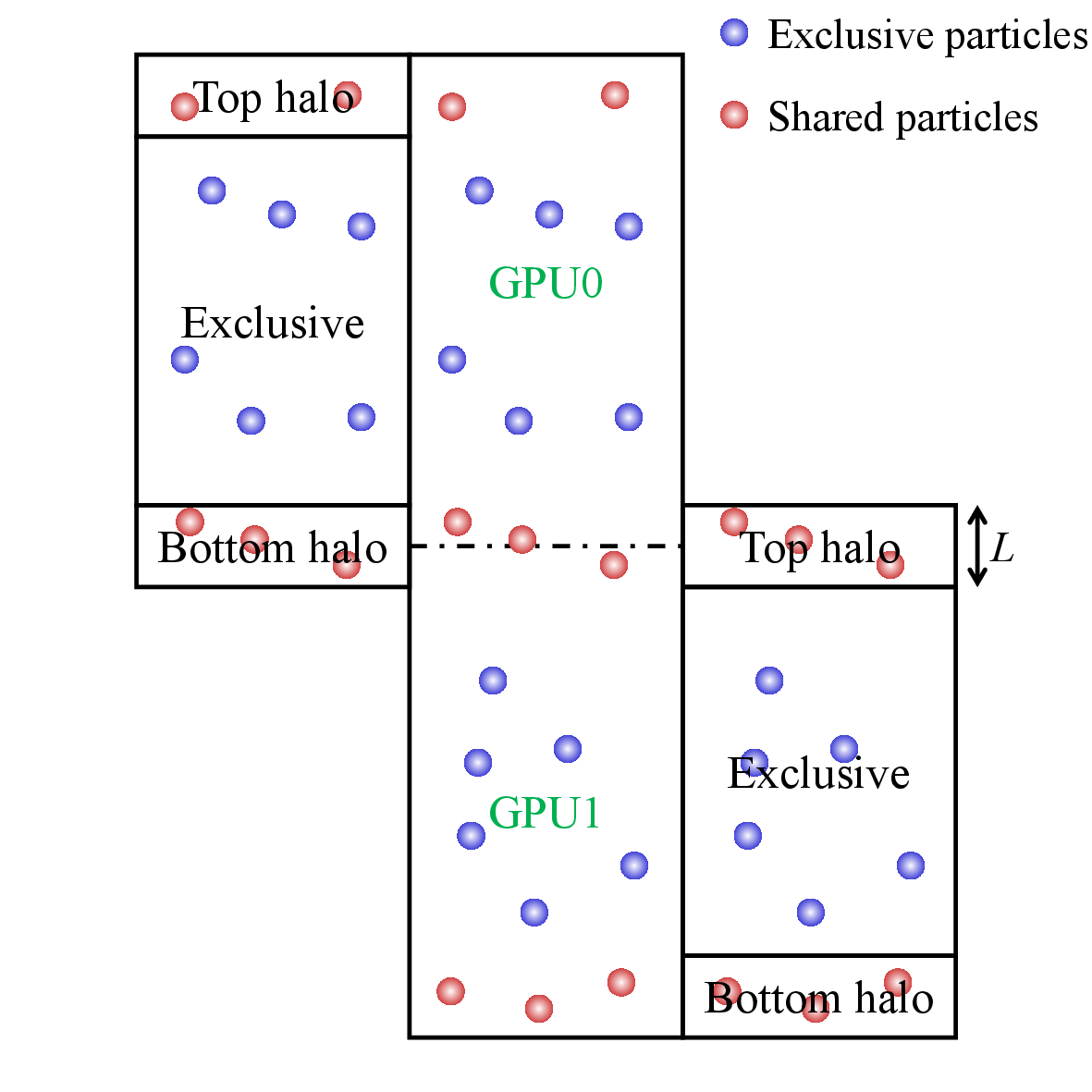} \\
  \caption{Schematic illustration to decompose the particle groups and build the domain list.}\label{Figure_domainListDemo}
\end{figure}

In the previous subsection, we decomposed the flow field into slices and assigned each subdomain to one GPU.
In this subsection, we further decompose the particles into subgroups and assign each subgroup to one GPU.
In the particle-resolved LB method, the fluid-particle interaction within a timestep only affects fluid nodes near the particle surface.
Thus, most of the particles in a subdomain do not have to share information with other subdomains, except for particles near the subdomain boundary.
As illustrated in Fig. \ref{Figure_domainListDemo}, we define an extended subdomain for a particle group compared to that for the flow field, which includes three parts:
the upper and lower parts with a length of $L$ (referred to as the top halo region and bottom halo region, respectively), and the central part similar as that of the sliced flow field (referred to as the exclusive region).
Particles whose centers in the halo region share their information with the neighboring subdomains; while particles in the exclusive region do not communicate their information with other subdomains.
Here, $L$ is determined based on the farthest distance a particle can interact with the fluid node and another particle.
The vertical length of the particle subdomain extends upward and downward by $d_{p} + s$, where $d_{p}$ denotes the particle diameter and $s$ denotes the threshold value for calculating the interparticle interaction force.
Thus, the shared region has a length of $L = 2(d_{p} + s)$.

After decomposing the particle groups, we assign each subgroup to a GPU and build domain lists to store the region index of each particle.
In this way, the particle-related calculations do not require iteration over all particles but only over the particles within the same list, thus reducing the size of the particle search.
To synchronize the forces of the particles, only particles in the top halo region and bottom halo region need to be shared with the adjacent subdomains.
Because the forces of the particles in the exclusive region do not require synchronization with other subdomains, we can reduce the size of messages that need to be sent.
If a particle enters the exclusive region from the top (or bottom) halo region, it will disappear from the subdomain list of the neighboring subdomains.
On the other hand, if a particle enters the top (or bottom) halo region from the exclusive region, information about this particle must be sent to the adjacent subdomain and added to their lists.
However, updating the domain list introduces additional computational costs, which may not be worth it for particles with low-volume fractions.

\begin{figure}
  \centering
  \includegraphics[width=\textwidth]{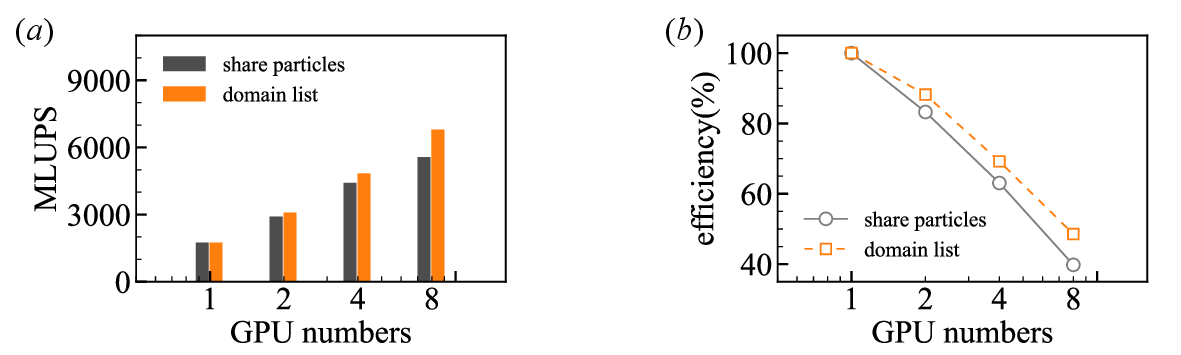}\\
  \caption{Performance comparisons between sharing particles' information among all GPUs and building domain list in terms of
  (\textit{a}) the MLUPS and (\textit{b}) the parallel efficiency.}\label{Figure_compare_domain-list}
\end{figure}

Fig. \ref{Figure_compare_domain-list} compares the performance of sharing particles’ information among all GPUs and building the domain list.
We can see that building domain list can improve parallel performance for all the cases.
Specifically, using 8 GPUs, the simulation can achieve 6796 MLUPS with a parallel efficiency of 48.5\%.
It should be noted that the idea of building domain list was inspired by the cell-linked list in smoothed particle hydrodynamics (SPH) simulations \cite{dominguez2011neighbour}, where the number of particles can range from hundreds of thousands to millions, leading to a heavy computational load when searching neighboring particles and calculating particle interaction forces.
To overcome this challenge, the simulation domain was divided into cells, and neighbor lists were created within each cell to limit the calculations to the same or adjacent cells.
In the particle-resolved LB method, the particle size is much larger than the grid size, and the particle number is much smaller than that in SPH.
Thus, it is unnecessary to further divide the subdomain assigned to each GPU into smaller ones;
in other words, we only build one domain list within a subdomain.
As the number of particles increases, the benefits gained from building domain lists will become more obvious.

\subsection{Utilizing request-answer communication}

\begin{figure}
  \centering
  \includegraphics[width=0.9\textwidth]{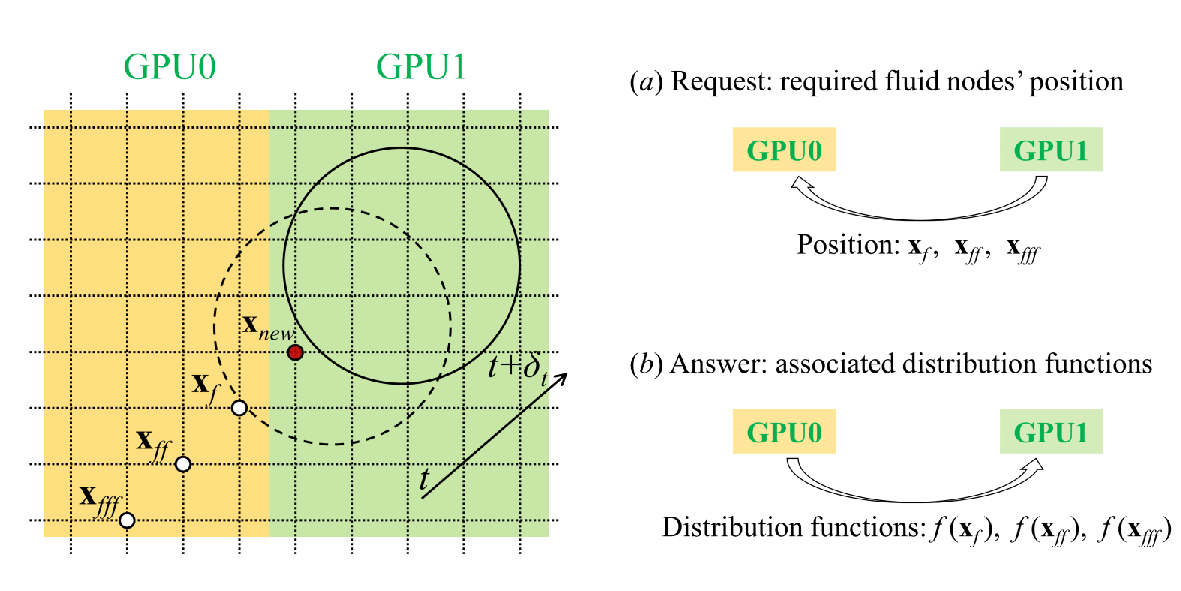}\\
  \caption{Schematic illustration of request-answer communication:
  (\textit{a}) sending the request to neighboring GPUs and (\textit{b}) returning the requested distribution function.}\label{Figure_requestAnswerDemo}
\end{figure}

The refilling scheme used to construct unknown distribution functions (as described in subsection \ref{sec:refilling}) can result in communication overhead if node information is required from neighboring subdomains.
Due to the constant change in particle position, it is difficult to predict in advance which node information needs to be communicated.
A straightforward solution is to exchange all possible information for extrapolation;
however, this approach degrades the utilization efficiency of exchanged data, particularly for particles with a large specific area.
To avoid the passing of excess data, we adopt the request-answer communication method between GPUs, which is similar to the communication method between CPUs \cite{geneva2017scalable}.
The basic idea behind this method is to exchange only distribution functions that are needed for the extrapolation, rather than exchanging all distribution functions at boundary nodes of nearby subdomains.
As illustrated in Fig. \ref{Figure_requestAnswerDemo}, when the particle surface moves from the dashed curve to the solid curve, the node $\mathbf{x}_{new}$ changes from a solid node to a fluid node.
We assume that the extrapolation of the density distribution function requires information on the fluid nodes $\mathbf{x}_{f}$, $\mathbf{x}_{ff}$, $\mathbf{x}_{fff}$, which are stored in the adjacent subdomain.
To exchange this information, GPU1 sends a request to GPU0 to exchange the information at fluid nodes $\mathbf{x}_{f}$, $\mathbf{x}_{ff}$, and $\mathbf{x}_{fff}$.
Upon receiving the request, GPU0 sends the information of distribution functions $f_{0-8}(\mathbf{x}_{f}, \delta_{t})$, $f_{0-8}(\mathbf{x}_{ff}, \delta_{t})$, and  $f_{0-8}(\mathbf{x}_{fff}, \delta_{t})$ to GPU1 in the order they were requested.
In this way, the fluid refilling calculation can be performed with only two communications.
Practically, we pack discontinuous data into contiguous memory, exchange the information using MPI, and then unpack the synchronized information to the desired location.
We allocate a sufficiently large array at the receiver to receive the entire message and use \emph{MPI GET COUNT} to obtain its length.
In the second communication, the distribution functions should be sent in the order that corresponds to the requested information, which makes it easier for the receiver to unpack the information.

\begin{figure}
  \centering
  \includegraphics[width=\textwidth]{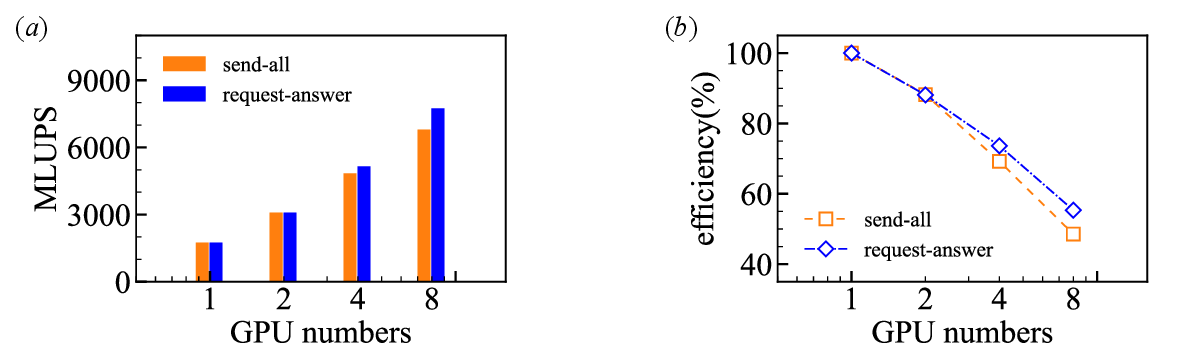} \\
  \caption{Performance comparisons between exchanging all boundary node information and utilizing request-answer communication in terms of
  (\textit{a}) the MLUPS and (\textit{b}) the parallel efficiency.}\label{Figure_compare_requestAnswer}
\end{figure}

Fig. \ref{Figure_compare_requestAnswer} compares the performance of the-above mentioned two approaches:
exchanging all boundary node information and utilizing the request-answer communication approach.
We can see the simulation can achieve 7746 MLUPS using 8 GPUs with a parallel efficiency of 55.3\%.
However, with a smaller number of GPUs, the parallel efficiency does not improve obviously.
Upon further analysis of the time consumption using the NVIDIA Nsight Systems tool, we find that even though the request-answer communication approach reduced the amount of data transfer between GPUs, the total time consumption (including both communication initiation and data transfer duration) increased.
This increase was likely due to the increased latency on the CPU side, resulting from multiple communication initiations and calls to the \emph{MPI GET COUNT} function.
When dealing with 8 or more GPUs, inter-GPU communication encounters limitations within the PCI-e interface, leading to a decrease in data transfer speeds and an increase in data transfer duration.
This is the scenario where the request-answer communication approach enhances parallel performance and overall efficiency.
In the 2D simulation, due to low communication overhead, the increased latency on the CPU side outweighs the benefit of reduced communications between GPUs.
For this reason, we do not use the request-answer approach to optimize the interpolation of the distribution function at the curved surface of the particle, as the additional communication load from the low percentage of invalid information was acceptable for high-volume fractions of particles.
We deduce that in three-dimensional simulations when the overhead between GPU communications is more intense, the use of a request-answer communication approach would be more effective.

\subsection{Overlapping communications with computations}

\begin{figure}
  \centering
  \includegraphics[width=\textwidth]{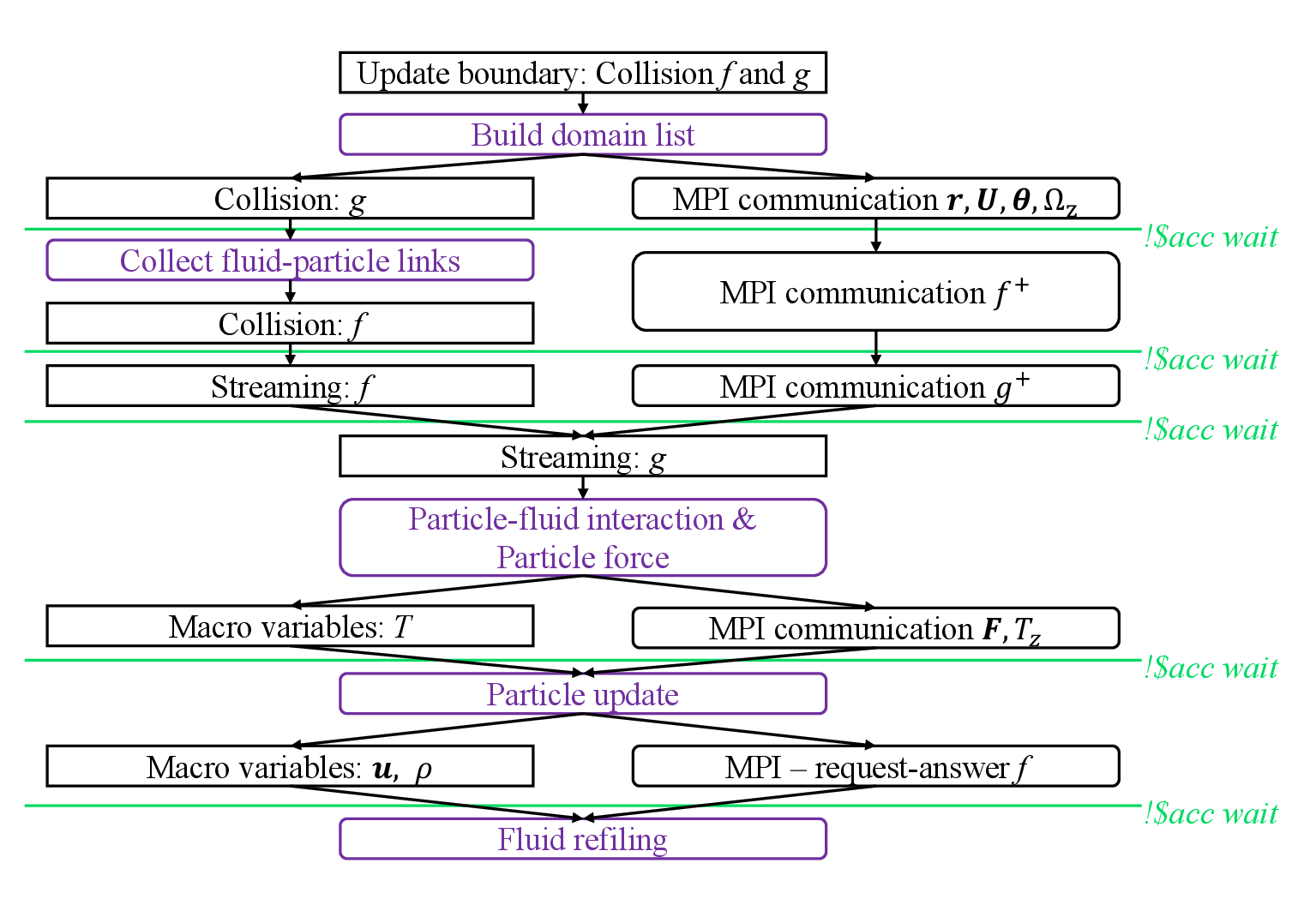}\\
  \caption{Schematic illustration of overlapping communications with computations for LB simulation of particle-laden thermal flows.
  The purple rectangular represents particle-related computations.}\label{Figure_overlapDemo}
\end{figure}

Previous studies have demonstrated that hiding communication overhead behind the kernel runtime can increase the parallel performance by a factor of around 1.3X for LB simulation of single-phase flow \cite{xu2023multi,xian2011multi,hong2015scalable}.
In Fig. \ref{Figure_overlapDemo}, we further illustrate the overlap of communications with computations for LB simulation of particle-laden thermal flows.
Specifically, we first carry out the collision step to update the density distribution function ($f$) and temperature distribution function ($g$) at boundary nodes, followed by building domain lists in each subdomain.
If a particle in a subdomain transition from an exclusive state to a shared state, namely it moves from the exclusive region to a top (or bottom) halo region, we use MPI communications to synchronize the particles’ information, including the position ($\mathbf{r}$), the velocity ($\mathbf{U}$), the orientation angle ($\theta$), and the angular velocity ($\Omega_{z}$).
The synchronization of the particle's information can be overlapped with updating the temperature distribution function ($g$) at inner nodes.
After that, we collect the particle-fluid grid linkage and carry the collision step to update the density distribution function ($f$) at inner nodes;
meanwhile, we synchronize the post-collision density distribution function $f^{+}$ to hide the kernel runtime.
Similarly, the synchronization of post-collision temperature distribution function $g^{+}$ can be overlapped with the streaming of density distribution function ($f$).
After the streaming of the temperature distribution function ($g$) and the calculation of fluid-particle interactions, the particles in the top (or bottom) halo region of the subdomain lists need to exchange the information of fluid-particle interaction force ($\mathbf{F}$) and torque ($T_{z}$) with the neighboring subdomains, which can be overlapped with the computation of macroscopic temperature ($T$).
Before the fluid refilling calculation, we synchronize the density distribution function ($f$) using the request-answer communication approach, which can be overlapped with the computation of macroscopic velocity ($\mathbf{u}$) and density ($\rho$).

\begin{figure}
  \centering
  \includegraphics[width=\textwidth]{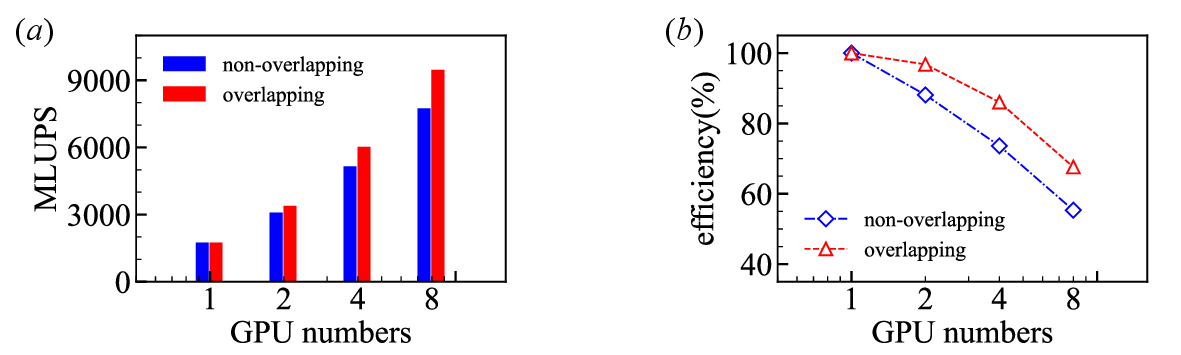}\\
  \caption{Performance comparisons between non-overlapping and overlapping communication with computation in terms of
  (\textit{a}) the MLUPS and (\textit{b}) the parallel efficiency.}\label{Figure_compare_overlapping}
\end{figure}

Fig. \ref{Figure_compare_overlapping} compares the performance between non-overlapping and overlapping communication.
We can see that the MLUPS and parallel efficiency improved for all the cases when the communications and computations overlapped, and the advantage of using the overlapping mode becomes more pronounced with an increase in the number of GPUs.
Notably, the performance increases by 1.22X when using 8 GPUs
(i.e., increase from 7746 MLUPS to 9466 MLUPS, and a parallel efficiency from 55.3\% to 67.6\%),
which shows similar gains to that for LB simulation of single-phase flow, suggesting that hiding communication overhead behind the kernel runtime is an effective approach for optimizing LB simulations on multi-GPUs \cite{xu2023multi,xian2011multi,hong2015scalable}.

\subsection{Executing computation tasks concurrently}

\begin{figure}
  \centering
  \includegraphics[width=\textwidth]{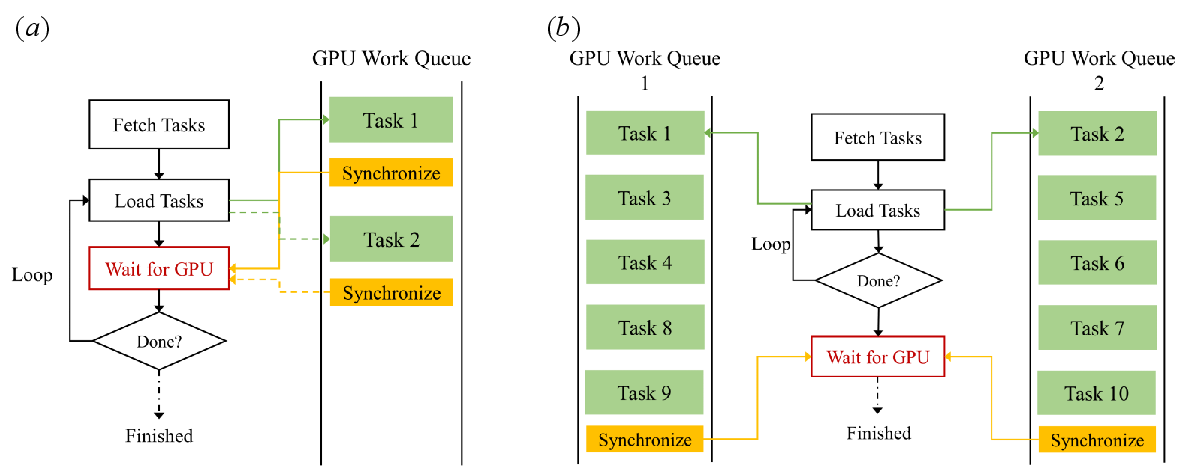}\\
  \caption{Schematic illustration of (\textit{a}) non-concurrent and (\textit{b}) concurrent execution of computational tasks on a single GPU.}\label{Figure_concurrencyDemo}
\end{figure}

In the heterogeneous CPU-GPU architecture, the CPU acts as the controller, which offloads data and computational tasks to the GPU, and retrieves the results when the computation is complete, as illustrated in Fig. \ref{Figure_concurrencyDemo}(a).
This non-concurrent computation can lead to performance bottlenecks if the CPU is unable to keep up with the demands of the GPU.
A solution is to take advantage of the parallel processing power of the GPU and execute computation tasks concurrently.
In the OpenACC, task parallelism can be achieved via the \emph{!\$acc async(n)} handle to concurrently execute independent tasks on a single GPU.
As illustrated in Fig. \ref{Figure_concurrencyDemo}(b), each task is processed by a separate stream of GPU commands, and the concurrent execution of computation tasks reduces synchronization between CPU and GPU.
By processing multiple tasks in parallel, the GPU can be kept busy, and utilization can be maximized.

\begin{figure}
  \centering
  \includegraphics[width=\textwidth]{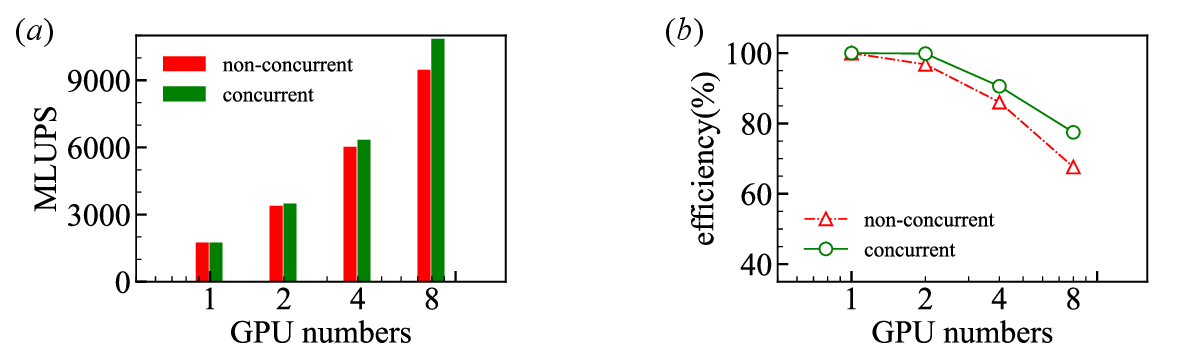}\\
  \caption{Performance comparisons between non-concurrent and concurrent computation in terms of
  (\textit{a}) the MLUPS and (\textit{b}) the parallel efficiency.}\label{Figure_compare_concurrent}
\end{figure}

Fig. \ref{Figure_compare_concurrent} compares the performance between non-concurrent and concurrent computation.
We can see that the MLUPS and parallel efficiency slightly improved for all the cases if the computations are executed concurrently.
With 8 GPUs, the simulation achieved 10846 MLUPS, yielding a parallel efficiency of 77.5\%.
We further analyzed the runtime of concurrent computation using the NVIDIA Nsight Systems tool and found that each GPU has three distinct work queues that simultaneously execute the computational tasks.

\section{Conclusions \label{sec:conclusions}}

In this work, we have utilized OpenACC-based GPU computing to perform particle-resolved thermal LB simulations, in which the momentum-exchange method was adopted to calculate particle-fluid interactions.
We extended the indirect addressing method to collect fluid-particle link information at each timestep and store indices of fluid-particle links in a fixed index array.
This mapping of the index array helps solve the issue of load imbalance by ensuring fluid-particle interactions are only calculated at indexed positions. Using this approach, the simulation of 4,800 hot particles settling in cold fluids with a domain size of $4000^{2}$ achieved 1750 MLUPS on a single GPU.

We also implemented a hybrid approach combining OpenACC and MPI for multi-GPU accelerated simulation.
This approach incorporates four optimization strategies to enhance parallel performance.
First, we build the domain list and optimize the fluid-particle interactions by considering only those within the same domain or adjacent domains, thereby avoiding the need to loop over all particles.
Next, we utilize request-answer communication and exchange only the necessary distribution functions, rather than exchanging information for all boundary nodes.
To further improve performance, we overlap communications with computations.
This allows us to hide communication latency behind the consumed computational time, resulting in significant gains for multi-GPU simulations.
Additionally, we maximize the utilization of GPU resources by executing computational tasks concurrently, enhancing parallel efficiency by ensuring efficient use of available processing power.
Overall, using 8 GPUs, these optimizations lead to a parallel performance increase from 5572 MLUPS to 10846 MLUPS, with a corresponding improvement in parallel efficiency from 39.8\% to 77.5\%.
To ensure the correctness of the code utilizing the hybrid OpenACC and MPI approach, we recommend an incremental approach utilize the above four optimization strategies to accelerate the code.

In the future, we plan to extend these optimization strategies to three-dimensional particle-resolved thermal flows, where the computational load is intense and the overhead to lunch the kernel is relatively lower.
In the three-dimension simulation, to reduce memory requirements, the distribution function can be reconstructed from available hydrodynamic variables instead of storing the full set of discrete populations  \cite{tiribocchi2023lightweight}.

\section*{Acknowledgements}

This work was supported by the National Natural Science Foundation of China (NSFC) through Grant Nos. 12272311 and 11902268, and the Open Fund of Key Laboratory of Icing and Anti/De-icing (Grant No. IADL20200301).
The authors acknowledge the Beijing Beilong Super Cloud Computing Co., Ltd for providing HPC resources that have contributed to the research results reported within this paper (URL: http://www.blsc.cn/).

\section*{Appendix A. An elliptical cold particle settling in hot fluids \label{sec:appendixElliptical}}

\begin{figure}
  \centering
  \includegraphics[width=0.8\textwidth]{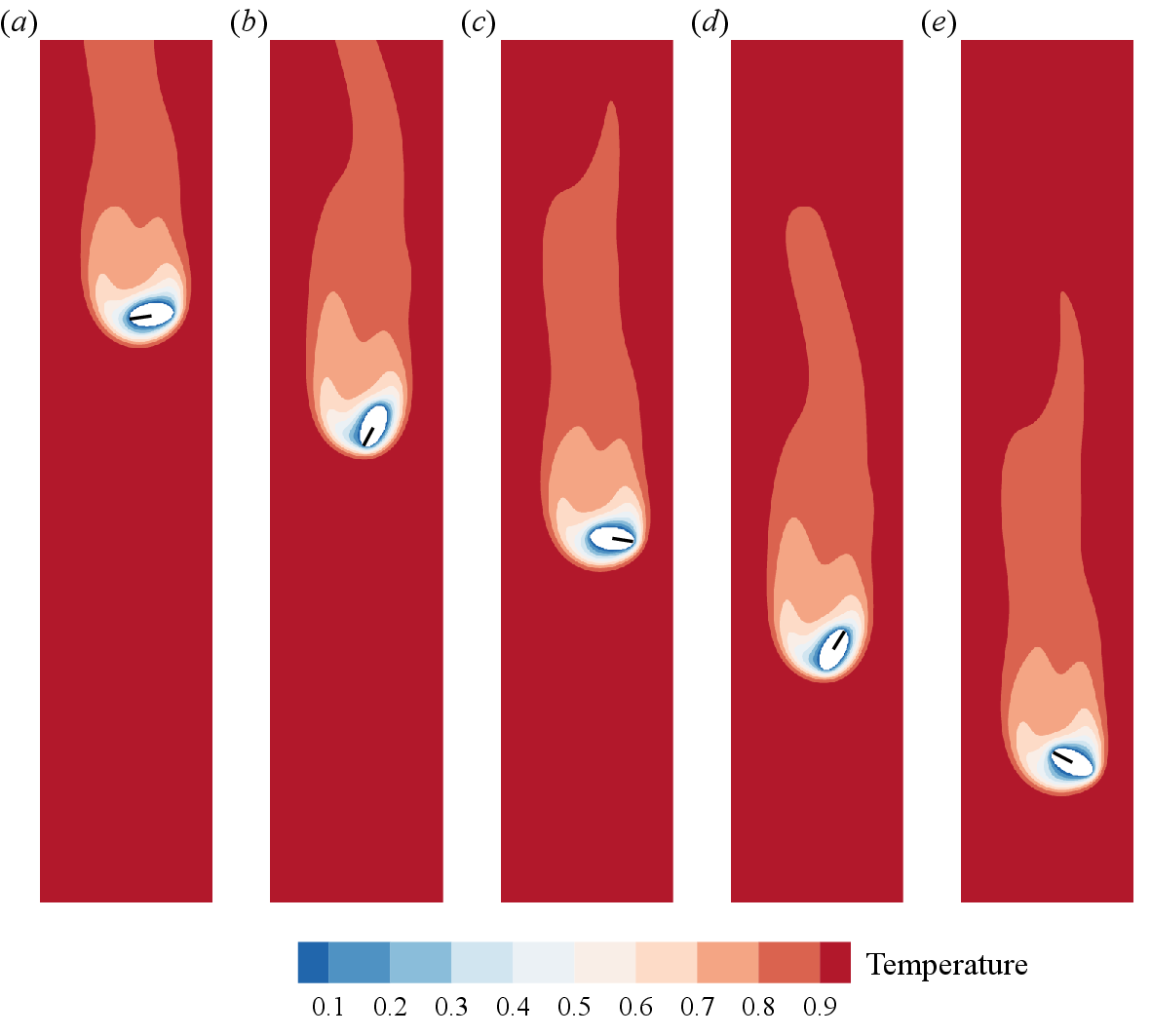}\\
  \caption{Contour of dimensionless temperature field $T^{*}=(T-T_{c})/\Delta_{T}$ during the sedimentation of an elliptical cold particle in a hot fluid at the dimensionless time $t^{*}=t\nu/A^{2}$  of
  (\textit{a}) 6.875, (\textit{b}) 7.5, (\textit{c}) 8.125, (\textit{d}) 8.75, (\textit{e}) 9.375.
  Note that only heights between 1.875 cm and 3.875 cm are shown for better visualization.}\label{Figure_singleElliptic-temperature}
\end{figure}

We compare the trajectory and orientation of a single elliptical cold particle settling in hot fluids using both the multi-GPU code and the corresponding CPU code.
The contour of the temperature field during the sedimentation is shown in Fig. \ref{Figure_singleElliptic-temperature}.
The simulation setting is similar to our previous work \cite{xu2018thermal}.
However, instead of using the moving domain technique to mimic an infinitely long channel, we now adopt an alternative approach of utilizing a closed cavity with a very small width-to-height aspect ratio \cite{suzuki2020numerical} to minimize the end effect of top and bottom boundaries.
Specifically, the size of the cavity is $W \times H$ = 0.4 cm $\times$ 8 cm, and all four walls of the cavity are imposed no-slip velocity boundary conditions.

\begin{figure}
  \centering
  \includegraphics[width=0.8\textwidth]{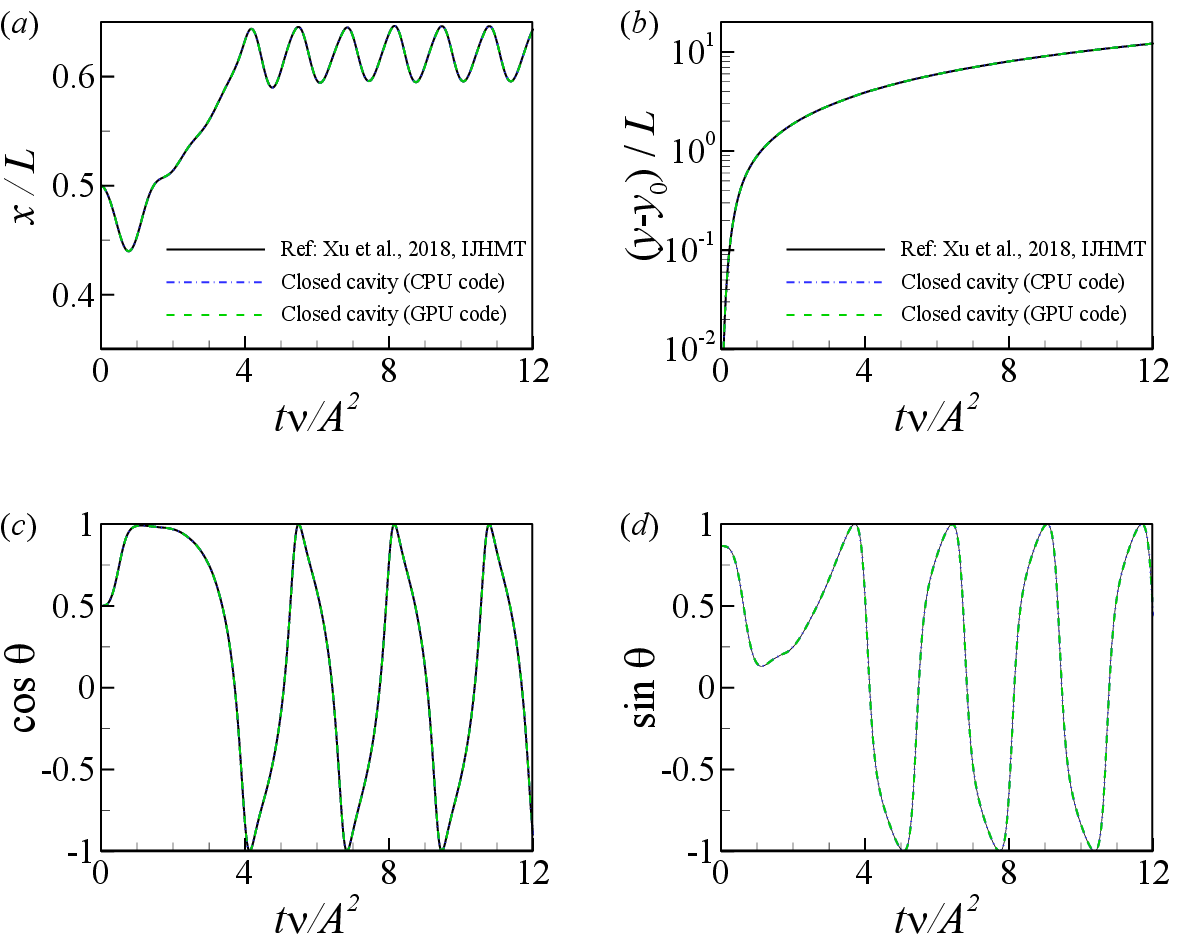}\\
  \caption{Time series of (\textit{a}, \textit{b}) horizontal and vertical positions of the particle center,
  and (\textit{b}, \textit{c}) the angle $\theta$ between the particle major axis and the horizontal direction in terms of $\cos \theta$ and $\sin \theta$, respectively.}\label{Figure_singleElliptic}
\end{figure}

In the simulation, each particle has a density of  $\rho_{p}$ = 1.001 g/cm$^{3}$, a major axis of $A$ = 1 mm, and a minor axis of $B$ = 0.5 mm.
The particle is released at $(0.5W,0.75H)$ with an initial angle of 60$^{\circ}$ between the particle's major axis and the horizontal direction.
To mimic the working fluid of water, we set its viscosity as $\nu_{f} = 10^{-6}$ m$^{2}$/s and its density as $\rho_{f}=1$ g/cm$^{3}$.
In this case, we have   $Re_{p}=U_{ref}d_{p}/\nu_{f}= 3.92$ and   $Ar=\sqrt{gA^{3}(\rho_{p}-\rho_{f})/(\nu^{2}_{f}\rho_{f})}= 3.13$.
Meanwhile, we choose $Gr_{p} = g\beta_{T}\Delta_{T}A^{3}/\nu_{f}^{2}  = 200$ and $Pr = 7$.
From Fig. \ref{Figure_singleElliptic}, we can see that the simulation using the moving domain technique  \cite{xu2018thermal} gives the same results as that adopting a sufficiently large domain;
in addition, the multi-GPU simulation provides the same results to those obtained from the CPU-based simulation.

\section*{Appendix B. Two circular hot particles settling in cold fluids \label{sec:appendixThermalDKT}}

\begin{figure}
  \centering
  \includegraphics[width=0.8\textwidth]{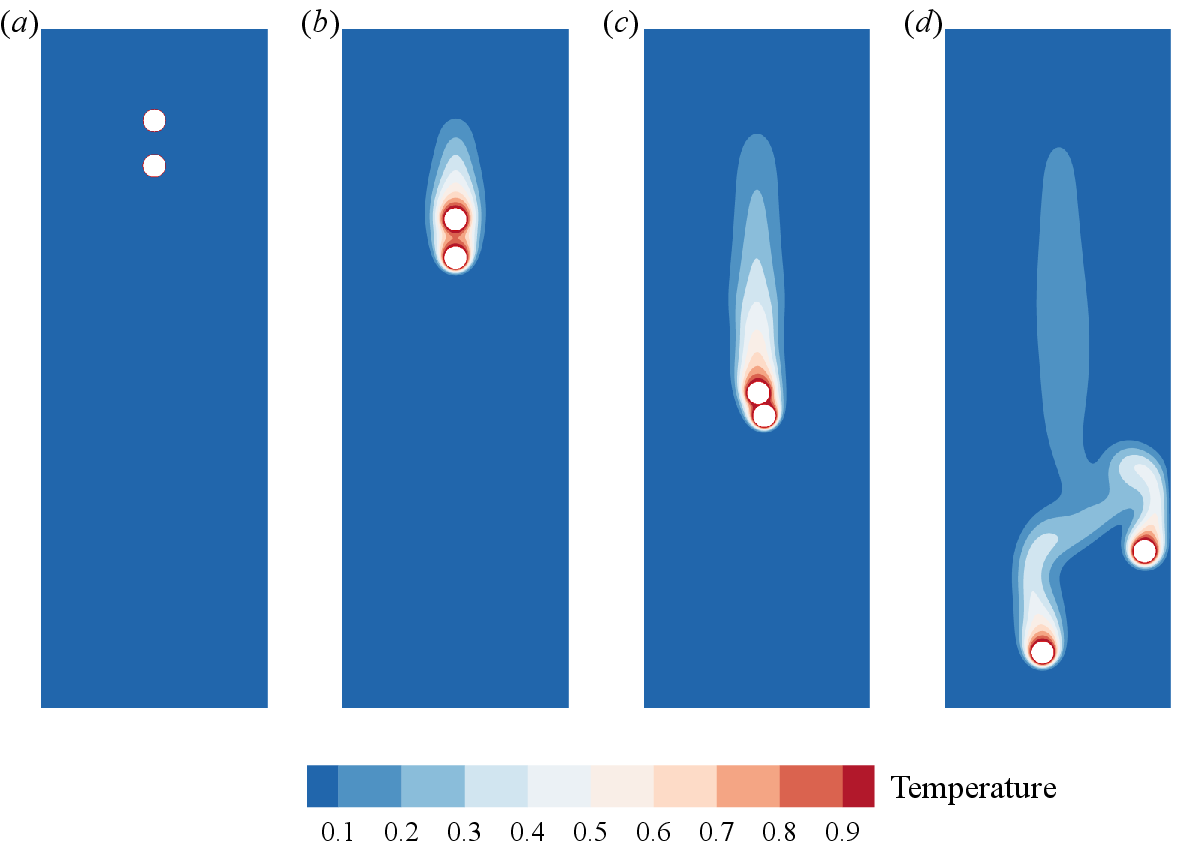}\\
  \caption{Contour of dimensionless temperature field $T^{*}=(T-T_{c})/\Delta_{T}$  during the sedimentation of two circular hot particles settling in cold fluids at the dimensionless time $t^{*}=t/\sqrt{d_{p}/g}$ of
  (\textit{a}) 0, (\textit{b}) 70, (\textit{c}) 140, (\textit{d}) 280. }\label{Figure_thermalDKT_temperature}
\end{figure}

We conduct a comparison between the velocity and position of two circular hot particles settling in cold fluids to validate the particle-particle interactions, and the contour of the temperature field during the sedimentation is shown in Fig. \ref{Figure_thermalDKT_temperature}.
The simulation setting is similar to the one used in Tao \emph{et al.} \cite{tao2022sharp}, known as draft-kissing-tumbling (DKT) with convection.
Specifically, the size of the cavity is $W \times H$ = 2 cm $\times$ 6 cm, and all four walls of the cavity are imposed no-slip velocity boundary conditions.
Each particle has a density of  $\rho_{p}$ = 1.01 g/cm$^{3}$ and a diameter of $d_{p}$ = 2 mm.
The lower particle is released at $(0.5W-0.005d_{p}, 0.8H)$ and the upper particle is released at $(0.5W, 0.8H+2d_{p})$.
The lower particle was deliberately offset from channel centerline to induce tumbling, because previous calculations showed that otherwise both particles remained perfectly aligned for a long time after catching up, and the offset prevents prolonged stable alignment.

\begin{figure}
  \centering
  \includegraphics[width=0.9\textwidth]{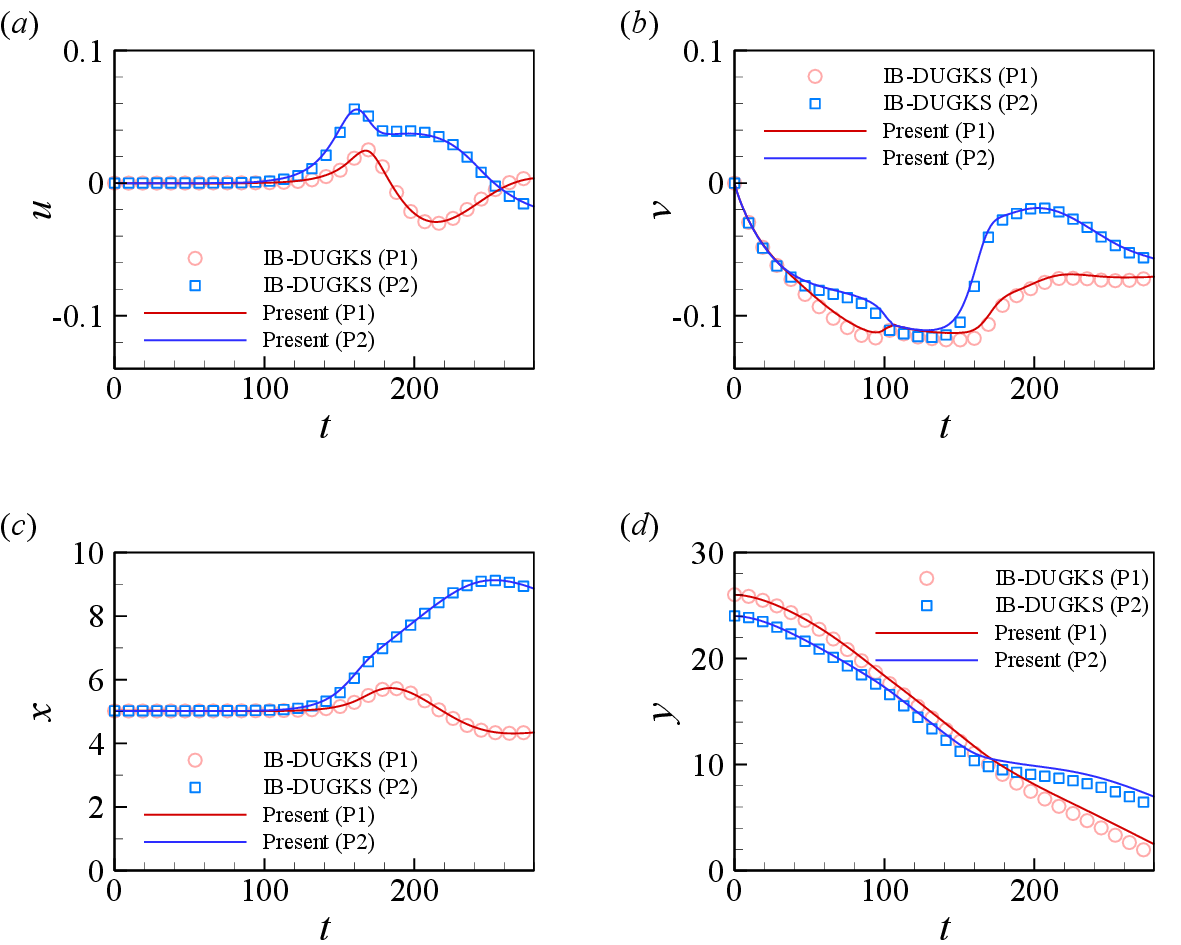}\\
  \caption{Time series of (\textit{a}, \textit{b}) velocity and (\textit{c}, \textit{d}) position of the two circular hot particles settling in cold fluids.
  The time is normalized as  $t^{*}=t/\sqrt{d_{p}/g}$, the velocity is normalized as  $\mathbf{u}^{*}=\mathbf{u}/\sqrt{gd_{p}}$, and the position is normalized as  $\mathbf{x}^{*}=\mathbf{x}/d_{p}$.
  Here, “P1” denotes the upper particle, and “P2” denotes the lower particle.
  Data shown in (\textit{b}) are from Tao \emph{et al.} \cite{tao2022sharp}, while data shown in (\textit{a},\textit{c},\textit{d}) are from private communication with S. Tao.}\label{Figure_thermalDKT}
\end{figure}

We set the viscosity of the fluid as $\nu_{f} = 10^{-6}$ m$^{2}$/s and its density as  $\rho_{f}=1$ g/cm$^{3}$.
In this case, we have  $Re_{p}=U_{ref}d_{p}/\nu_{f}=\sqrt{g\pi d_{p}(\rho_{p}-\rho_{f})/(2\rho_{f})}= 35.09$ and $Ar=\sqrt{gd_{p}^{3}(\rho_{p}-\rho_{f})/(\nu^{2}_{f}\rho_{f})} = 28.0$.
Meanwhile, we choose $Gr_{p} = g\beta_{T}\Delta_{T}A^{3}/\nu_{f}^{2}  = 100$ and $Pr = 1$.
A detailed setting for simulation parameters is listed in Table \ref{tb:thermalDKT}.
From Fig. \ref{Figure_thermalDKT}, we can see that the present results show good agreement with Tao \emph{et al.} \cite{tao2022sharp} using a sharp interface immersed boundary-discrete unified gas kinetic scheme (IB-DUGKS), thus validating the code's ability to simulate particle-particle interactions.

\begin{table}
\centering
\caption{Simulation parameters for the sedimentation of two circular hot particles in cold fluids.
The length unit conversion is $l_{*} = 5\times 10^{-5}$ m/l.u.,
the time unit conversion is $t_{*} = 2.5\times 10^{-5}$ s/t.s.,
and the temperature unit conversion is $T_{*} = 6.07$ K/t.u.}
\begin{adjustbox}{center, max width=\textwidth}
\small
\begin{tabular}{cccc}
  \hline
              & Physical system & LB system & Unit conversion \\
  \hline
  Domain size & $W\times H=$ 2 cm $\times$ 6 cm & $\overline{W} \times \overline{H}=$ 400 l.u. $\times$ 1200 l.u. & $\mathbf{x}=\overline{\mathbf{x}}\cdot l_{*}$ \\
  Particle diameter    & $d_{p}=2$ mm                       & $\overline{d_{p}}=40$ l.u. & $d_{p}=\overline{d_{p}}\cdot l_{*}$ \\
  Kinematic viscosity  & $\nu_{f} = 10^{-6}$ m$^{2}$/s	    & $\overline{\nu_{f}}$ = 0.01 l.u.$^{2}$/t.s. & $\nu_{f}=\overline{\nu_{f}}\cdot l_{*}^{2}/t_{*}$ \\
  Thermal diffusivity  & $\alpha_{T}= 10^{-6}$ m$^{2}$/s    & $\overline{\alpha_{T}}$ = 0.01 l.u.$^{2}$/t.s. & $\alpha_{T}=\overline{\alpha_{T}}\cdot l_{*}^{2}/t_{*}$ \\
  Gravity acceleration & $g = 9.8$ m/s$^{2}$	            & $\overline{g}= 1.23 \times 10^{-4}$ l.u./t.s.$^{2}$  & $g=\overline{g}\cdot l_{*}/t_{*}^{2}$ \\
  Thermal expansion coefficient	& 	 $ \beta = 2.1\times 10^{-4}$ K$^{-1}$  &  $\overline{\beta} = 1.28\times 10^{-3}$ t.u.$^{-1}$ & $\beta=\overline{\beta}/T_{*}$ \\
  Temperature difference	& 	 $\Delta_{T} = 6.07$ K  &  $\overline{\Delta_{T}}$ = 1 t.u. & $\Delta_{T}=\overline{\Delta_{T}}\cdot T_{*}$ \\
  \hline
\end{tabular} \label{tb:thermalDKT}
\end{adjustbox}
\end{table}

\section*{References}


\end{document}